\documentstyle[aps,prl,twocolumn,epsfig]{revtex}

\begin{document}

   \draft

   %Fixing abstract in twocolumn mode

   \twocolumn[\hsize\textwidth\columnwidth\hsize\csname @twocolumnfalse\endcsname%

   %\twocolumn

   \title{H\"older exponent spectra for human gait}
   \author{N. Scafetta$^{1}$,  L.  Griffin$^{1}$ and B. J. West$^{1,2}$.}
   \address{$^{1}$ Pratt School EE Dept., Duke University,  P.O. Box 90291, Durham, NC 27708. }
   \address{$^{2}$ Mathematics Division, Army Research Office, Research Triangle Park, NC. }
    \date{\today}
   \maketitle

   \begin{abstract}
The stride interval time series in normal human gait is not strictly constant, but
fluctuates from step to step in a complex manner. More precisely, it has been shown that  the control process for human gait is a fractal
random phenomenon, that is, one with a long-term memory. Herein we study the H\"older exponent spectra for the slow, normal and fast  gaits of 10 young healthy men in both free and metronomically triggered conditions and establish that the stride interval time series
is  more complex than a monofractal phenomenon.  A slightly multifractal and non-stationary time series under the three different gait conditions emerges. 

 \end{abstract}
   \vspace{0.5cm}
   %
   %Fixing abstract in twocolumn mode%%%
   ]
   %%%%
   %

%%%%%%%%%
%%%%%%%%%
\section{Introduction}
In the past two decades we have witnessed an explosion in the biophysics and
physiological literature with regard to the identification of phenomena having long-term memory and
probability densities that extend far beyond the typical tail region of
Gaussian distributions. One way these processes have been classified is as
1/f-phenomena, since their time series have spectra that are inverse power
law in frequency or their probabilities are inverse power law. In either case the underlying structure is fractal, either in
space, time or both \cite{bass94}. Herein we are interested in
demonstrating that human gait time series, see Fig. 1,  is more than   monofractal. By estimating the H\"older exponents and their  spectra  using wavelet transform \cite{struzik}, we show that the stride interval time series is
weakly  multifractal;  the time series is sometimes non-stationary  and its fractal variability  changes in different  gait mode regimes.

The gait data we study are in public domain archives Physionet \cite{physionet}, which we downloaded. The data sets are the stride interval time series for 10 healthy young men walking at slow, normal and fast paces in both free and metronomically triggered conditions for a period of one hour in the former and 30 minutes in the latter cases, respectively. These data were originally collected and used by Hausdorff et al. \cite{Hausdorff4} to determine the dependence of the fractal dimension of the time series on  changes of the  average rate of walking. Given their positive results it is not a surprise that the gait time series is not monofractal, but is multifractal, with a dependence on the average  rate of walking. 

In Sec. 2 we give a short introduction to the phenomenon of locomotion, the traditional methods for modeling using the Central Pattern Generator (CPG), and review the data processing used to establish the fractal behavior of stride time interval time series. Sec. 3 reviews various ways to estimate the H\"older exponents and singularity  spectra.
In particular, the relationship between the singularity spectrum and the probability density for the realization of a particular H\"older exponent is discussed in the context of modeling the motocontrol system. The results of the data processing are presented in Sec. 4. Finally, in Sec. 5 we discuss  our results and compare them with those of Hausdorff et al. \cite{Hausdorff4}.

%%%%%
\section{Complexity of  locomotion}

Walking consists of a sequence of steps and the corresponding time series consists of the time intervals for these steps.
These steps may be partitioned into two phases: a stance phase and a swing
phase. The stance phase is initiated when a foot strikes the ground and ends
when it is lifted. The swing phase is initiated when the foot is lifted and
ends when it strikes the ground again. The time to complete each phase
varies with the stepping speed.  A stride interval is the length
of time from the start of one stance phase to the start of the next stance
phase. It is the variability in the time series made from these stride intervals that we address in this paper.

Traditionally the  legged locomotion of animals is understood through the
use of a Central Pattern Generator (CPG), an intraspinal network of neurons
capable of producing a syncopated output \cite{collins1}. The implicit
assumption in such an interpretation is that a given limb moves in direct
proportion to the voltage generated in a specific part of the CPG.
Experiments establishing the existence of a CPG have been done on animals
with spinal cord transections. It has been shown that such animals are
capable of walking under certain circumstances. Walking, for example, in a
mesencephalic cat, a cat with its brain stem sectioned rostral to the
superior colliculus, is very close to normal, on a flat, horizontal surface,
when a section of the midbrain is electrically stimulated. Stepping
continues as long as a train of electrical pulses is used to drive the stepping. This
is not a simple linear response process, however, since the frequency of the
stepping increases in proportion to the amplitude of the stimulation,
whereas changing the frequency of the driver has little effect of the
walking \cite{mann}. 

As Collins and Richmond \cite{collins1} point out, in spite of the studies
establishing the existence of a CPG in the central nervous system of
quadrupeds, such direct evidence does not exist for a vertebrate CPG for
legged locomotion. Consequently, these and other authors have turned to the
construction of models, based on the coupling of nonlinear oscillators, to
establish that the mathematical models are sufficiently robust to mimic the
locomotion characteristics observed in the movements of segmented bipeds 
\cite{cohen}, as well as in quadrupeds \cite{collins2}. These
characteristics, such as the switching among multiple gait patterns, is
shown to not depend on the detailed dynamics of the constituent nonlinear
oscillators, nor on their inter-oscillator coupling strengths \cite{collins1}.

It has been known for over a century that there is a variation in the stride
interval of humans during walking of approximately 3-4\% \cite{vierordt}.
This random variability is so small that the biomechanical community has
historically considered these fluctuations to be an uncorrelated random
process. In practice this means that the fluctuations in gait were thought
not to contain any useful information about the underlying motocontrol
process. On the other hand, Hausdorff \textit{et al}. \cite
{Hausdorff4,hausdorff1} demonstrated that stride-interval time series
exhibit long-time correlations, and suggested that the phenomenon of walking
is a self-similar, fractal, activity. Subsequent studies by West and Griffin 
\cite{west1,west2,griffin} support these conclusions using a completely
different experimental protocol for generating the stride-interval time
series data and very different methods of analysis. The existence of fractal
time series suggests that the nonlinear oscillators needed to model
locomotion operates in the unstable, that is, in the chaotic regime.

A stochastic version of a CPG was developed by Hausdorff et al. \cite{Hausdorff4} to capture the fractal properties of the inter-stride interval time series. This stochastic model was later extended by Ashkenazy et al. \cite{ashkenazy} to describe the changing of gait dynamics as we develop from being a child to being an adult. The model is essentially a random walk on a chain, where each node of the chain is a neural center of the kind discussed above, and with a different frequency. This random walk is found to generate a multifractal rather than a monofractal process, with a width that depends parametrically on the range of the walker's step size. Ashkenazy et al. \cite{ashkenazy} focused on explaining the changes in the gait time series during maturation, using the stochastic CPG model. In addition, they applied the multifractal formalism to the computer-generated time series, to obtain the singularity spectrum, but they did not obtain such a spectrum using  the experimental  inter-stride interval time series. A related model, using a fractional Langevin equation, was proposed by West et al. \cite{bjwest} in which the multifractality of the signal is interpreted as a fluctuating scaling parameter.

Herein we use stride interval data to further refine the stochastic analysis for human gait by estimating the local H\"older exponents of the stride interval that measure the local fractality of a time series and the properties of their distributions.

%%%%%%
\section{H\"older exponent distribution}

In this section we give a short review of the analysis method  we  apply to  human gait data.   
In  {\it The Fractal Geometry of Nature} \cite{2Mandelbrot}, Mandelbrot showed that many natural phenomena are described by self-affine, correlated, time series. The scaling properties of such fractal noise, Fractional Gaussian Noises (FGN), are characterized by an exponent that Mandelbrot called $H$ in honor of Hurst.  If $X(t)$ is a fractal process with
Hurst exponent $H$ and $c$ is a constant, then $X\left( t\right) =X(ct)/c^{H}$ is another
fractal process with the same statistics. 
FGN is  defined by  a spectrum satisfying the power law
\begin{equation}\label{fbnps}
P(f) \propto f^{-\beta}~,
\end{equation}
where $f$ is the frequency, the exponent $\beta=(2H-1)$ and $H$ is the Hurst exponent. The self-affine property expressed by (\ref{fbnps}) and the relation between $\beta$ and $H$  are theoretically valid  only for an infinitely long monofractal noise.

Fig. 2 shows a computer generated realization of FGN with Hurst exponent $H=1$, also known as $1/f$ noise or {\it pink} noise. This type of noise is important because it is a kind of threshold between the persistent-stationary noise ($0.5<H<1$ or, assuming the asymptotic relation between the two parameters, $0<\beta<1$) and the non-stationary noise ($\beta>1$). Note that the common term {\it random} or {\it white noise} is characterized by $\beta=0$ or $H=0.5$ whereas the {\it random walk} or {\it Brownian motion} is characterized by $\beta=2$. The noise characterized by $0<H<0.5$ is called {\it anti-persistent}, whereas the noise characterized by $0.5<H<1$ is called {\it persistent} \cite{2Mandelbrot}. In the same way, we may call {\it anti-persistent} the walk given by the fluctuations characterized by $1<\beta<2$ and  {\it persistent} the walk given by the fluctuations characterized by $2<\beta<3$. The fluctuations characterized by an exponent $1<\beta<3$ we call  {\it walks} because they may be obtained by integrating fractal noises characterized by Hurst exponents in the interval  $0<H<1$. 

In our numerical example, we use  {\it pink} noise  because the gait data, as shown in the next section, are characterized by properties that range from a strong persistent-stationary noise to a weak non-stationary walk.  As seen in Fig. 2, a fractal noise is characterized by trends and discontinuities  that give a particular geometric shape to the signal.   The rapid changes in the time series are called {\it singularities} of the signal and their strength is measured by a  H\"older exponent  \cite{struzik}. Given a function $f(x)$ with a  singularity at $x_0$, the  H\"older exponent  $h(x_0)$ at such a point is defined as the supremum of all exponents $h$ that  fulfills the  condition:
\begin{equation}\label{holdedef}
|f(x)-P_n(x-x_0)|\leq C|x-x_0|^{h}~,
\end{equation}
where $P_n(x-x_0)$ is a polynomial of degree $n<h$. The relation between the H\"older and Hurst exponents in the continuum limit  is $h=H-1$.

\subsection{Continuous wavelet transform.}

The  H\"older exponent of a singularity can be evaluated by using the wavelet transform.
Wavelet analysis  \cite{daubechies,Mallat,percival} has become  a  powerful method to analyze time series.
Wavelet transforms makes use of scaling functions that have the 
property of being  localized in both time and frequency. 
A scaling coefficient $s$ characterizes and measures the width of a wavelet. 
 Given a signal $f(x)$, the continuous wavelet transform (CWT) of $f(x)$ is defined by
\begin{equation}\label{cwtdhjj3}
W_{s, x_0}(f)=\int\limits_{-\infty }^{\infty } \frac{1}{s}~ \psi\left(\frac{x-x_0}{s} \right)~f(x)~dx~,
\end{equation}
 where the kernel $\psi(u)$  is
 the wavelet filter centered at the origin, $u=0$,  with unit width, $s=1$. 

The wavelet transform can be used to determine the H\"older exponent of a singularity because the wavelet kernel $\psi(u)$ can be chosen in such a way as to be orthogonal to polynomials up to degree $n$, that is, such that the following properties are fulfilled
\begin{equation}\label{wavort}
\int\limits_{-\infty }^{+\infty } \frac{1}{s}~ \psi\left(\frac{x-x_0}{s} \right)~x^m~dx=0~~~\forall m, ~0\leq m \leq n~.
\end{equation}
In fact, if (\ref{wavort}) holds true, it is easy to prove that if the function $f(x)$ fulfils  condition (\ref{holdedef}), its wavelet trasform at $x=x_0$ is given by
\begin{equation}\label{wavcoeff}
W_{s, x_0}(f)=C |s|^{h(x_0)} \int\limits_{-\infty }^{+\infty }|u|^{h(x_0)}~\psi(u)~du \propto |s|^{h(x_0)}~,
\end{equation}
where $u=(x-x_0)/s$. Therefore, at least theoretically, the  H\"older exponent of a singularity can be evaluated as the scaling exponent of the wavelet transform coefficient, $W_{s, x_0}(f)$, for $s \to 0$.

 In this paper we make use of the Mexican Hat wavelet, the second derivative of a Gaussian,  and which  is defined by 
\begin{equation}\label{mexhat}
\psi(u) =(1-u^2) \exp(-u^2/2)~.
\end{equation}
The Mexican Hat wavelet integrates to zero polynomial biases up to degree $n=1$. Finally, due to the exponential convergence to zero of Eq. (\ref{mexhat}) for large $|u|$, we may assume that the Mexican  Hat wavelet explores a window size  approximately 10 times the scale $s$.

\subsection{Maxima lines and multifractal formalism.}

Even if Eq. (\ref{wavcoeff}) can  be evaluated for any position $x_0$, we are interested only in  the cusp singularities of the time series. Mallat  et al. \cite{Mallat,mallat2,mallat3} show that the H\"older exponent of these singularities can be evaluated by studying the scaling exponent $h(x_0)$ along the so-called {\it maxima line} that converges towards the singularity. The maxima lines are defined by the extremes of the wavelet transform coefficients (\ref{cwtdhjj3}) at each wavelet scale $s$. 

Fig. 3 shows the wavelet transform modulus maxima  (WTMM) line tree of the fractal noise of Fig. 2. Fig. 3 shows that, in a complex process, the singularities become less and less isolated as the scaling coefficient $s$ increases. This means that it is not possible, in general, to use  Eq. (\ref{wavcoeff}) to evaluate the H\"older exponent of a singularity because the maxima line of a singularity will be deformed by its neighboring singularities.  However,  Arneodo et. al. \cite{arneodo1,arneodo2} proved that WTMM can be used to  define a multifractal-like formalism that gives the stochastic properties of the singularities of a fractal or multifractal noise.

The idea is to evaluate the mass exponent $\tau(q)$ of the moment $q$ from the partition function $Z(s,q)$ assuming that
\begin{equation}\label{arzetaq}
Z(s,q)=\sum _{\Omega (s)} |W_{s, x_0}(f)|^q \propto s^{\tau(q)}~,
\end{equation}      
where the sum is over the set $\Omega (s)$ of all maxima at the scale $s$. The entire distribution of H\"older exponents can be determined because negative $q$ stress weak singularities  whereas positive $q$ stress strong singularities. Finally, by using the Legendre transformation it is possible to define the spectrum of the singularities $D(h)$ by
\begin{equation}\label{spesing}
{h(q)=d\tau(q)/dq \atop D[h(q)]=q h(q) - \tau(q)~.}
\end{equation}
Eq. (\ref{spesing}) gives a global average of the strength of the singularities of the time series and it has been recently used to determine the multifractal nature of many signals, for example,  that for  human heartbeats \cite{Stanley1}.

\subsection{Approximate estimation of local H\"older exponents and their probability distribution}

The spectrum of the singularities $D(h)$ defined by Eqs. (\ref{spesing}) presents some problems of stability when applied to observational data. In fact, the spectrum can be corrupted by the divergences of negative moments \cite{Mallat,arneodo2,struzik2} or by the {\it outliers}, that is, the end points of the sample singularities \cite{struzik}. Different methods have been suggested to remove the divergences due to the negative moments of the multifractal partition function, for example, by chaining the wavelet maxima across scales \cite{Mallat} or, more efficiently, by bounding the H\"older exponent of the maxima line by using the {\it slope} wavelet \cite{struzik2}.  The corruption of the singularity spectrum due to the outliers is more difficult to deal with.

Struzik \cite{struzik} suggested an alternative method that has the ability to determine an approximate value of local singularity strength. The spectrum may then be evaluated  from these approximate values.  
The idea is to estimate the mean H\"older exponent $\overline{h}$ as a linear fit of the following equation
\begin{equation}\label{linfitesa}
\log[M(s)]=\overline{h}~\log(s) +C~,
\end{equation}
where the function $M(s)$ is obtained via the partition function, Eq. (\ref{arzetaq}),
\begin{equation}\label{linfitesa2}
M(s)=\sqrt{\frac{Z(s,2)}{Z(s,0)}} ~.
\end{equation}
Eqs. (\ref{linfitesa}) and (\ref{linfitesa2}) allow us to consider the mean  H\"older exponent $\overline{h}$ to be the local version of the Hurst exponent $H$ \cite{2Mandelbrot}. More precisely, for a monofractal noise with Hurst exponent $H$, we have $\overline{h}=H-1$ because the Hurst exponent is evaluated by integrating the noise \cite{2Mandelbrot,feders}. Here again the equality is only rigorously true for an infinitely long monofractal noise data set.
The approximate local H\"older exponent $\hat{h}(x_0,s)$ at the singularity $x_0$ can now be evaluated as the slope
\begin{equation}\label{locholexp}
\hat{h}(x_0,s)=\frac{\log(|W_{s, x_0}(f)|)-(\overline{h}~\log(s) +C)}{\log(s)-\log(s_N)}~,
\end{equation}
where $s_N$ is the length of the entire wavelet maxima line tree, that is, the maximum available scale  that coincides with the sample length $s_N=N$, and $x_0$ belongs to the set $\Omega (s)$ of all wavelet maxima at the scale $s$ that assume the value $W_{s, x_0}(f)$.

Fig. 4 gives a graphical view of how the local H\"older exponent $\hat{h}(x_0,s)$ are evaluated. We plot $\log[|W_{s, x_0}(f)|]$ against $\log(s)$ for all wavelet maxima at many scales. The black crosses are  the mean H\"older exponents evaluated by using Eq. (\ref{linfitesa}) and are fitted in the interval [0,3] that corresponds to the scales $1 \leq s \leq 20$,  that yields $\overline{h}=-0.004 \pm 0.008$ and $C=1.00 \pm 0.02$. The intersection point on the right of the fitting line is the root of the wavelet line tree and corresponds to $\log(s_N)=\log(5000)=8.52$. Finally, the local H\"older exponent $\hat{h}(x_0,s)$ are evaluated as the slope of the straight line that joins the root of the wavelet line tree to the value of the wavelet coefficient $\log[|W_{s, x_0}(f)|]$ at each singularity at the scale $s$. The slope of the two straight lines shown in the figure represent the maximum and minimum value of $h$.

Fig. 5 shows the H\"older exponents for the FGN plotted against the position in the time series. Finally, Fig. 6 shows the histogram of H\"older exponents obtained using a computer-generated data set with a Hurst exponent $H=1$. The probability distribution, that gives the spectra of the H\"older exponents, is estimated with a histogram whose number of bins  is chosen, for stochastic stability, equal to the square root of the number singularities at the scale analyzed. In our example we use the smallest available scale, that is, $s=1$. Note that to obtain a shape similar to that usually seen for  a multifractal singularity spectrum, the probability distribution $p(h)$ should be graphed on a log-linear graph paper. In fact, according to the thermodynamic picture of multifractal behavior developed in Ref. \cite{beck}, the number $M(h)$ of boxes of size $\varepsilon$ with a H\"older exponent $h$ has the scaling behavior
\begin{equation}\label{boxbek}
M(h) \sim  \varepsilon ^{-D(h)}~,
\end{equation}
where $D(h)$ is the singularity spectrum. Introducing $V=-\log(\varepsilon)$, the probability that the H\"older exponent  in the interval $(h,h+dh)$ is
\begin{equation}\label{prohhbex}
p(h)=\frac{\exp[V~D(h)]}{Z}~,
\end{equation}
where $Z$ is the partition function. Eq. (\ref{prohhbex}) suggests that the singularity spectrum $D(h)$ can be interpreted as the entropy density of the escort distribution $S(h)=\log[p(h)]$ in the limit $\varepsilon \to 0$ limit, that is,
\begin{equation}\label{limback}
\lim _{V \to \infty } \frac{S(h)}{V} = D(h)~.
\end{equation}

\subsection{Fractal or multifractal?}

Fig. 6 shows the   histogram  of H\"older exponents for the artificial fractal noise of Fig. 2 produced with the Hurst exponent $H=1$.  Actually, we plot the probability density $p(h)$ against $h$, that is, the probability to find an H\"older exponent in a bin of size $\varepsilon$ divided by the size $\varepsilon$. The size $\varepsilon$ is determined by dividing the range between the maximum and the minimum of the H\"older exponents by the square root of the number af all  evaluated H\"older exponents.  The  probability distribution of H\"older exponents of our computer generated noise  has as its mean  $\overline{h}=-0.004 \pm 0.008$ that is consistent with the value $h=H-1=0$ that characterizes pink noise.  By fitting the histogram with the normalized Gaussian
\begin{equation}\label{gaussf}
g(h)=\frac{1}{\sqrt{2\pi} ~\sigma}~\exp{\left[-\frac{(h-h_0)^2}{2~\sigma^2} \right]}
\end{equation}
it is possible to evaluate the width of the distribution $\sigma$. Note that usually $h_0$ may be slightly larger than $\overline{h}$ because the distribution of the H\"older exponents may present a slightly positive skewness, so, in this computer generated noise, we measure $h_0=0.007 \pm 0.001$ and $\sigma=0.052\pm0.002$. The width $\sigma$ is not zero, as  would be expected for an infinitely long  computer-generated monofractal noise. This non-zero $\sigma$  may be mistaken for an indicator of  multifractal behavior. However, the non-vanishing value of $\sigma$ for our computer-generated noise  is due to the fact that a monofractal noise has some variability of the local H\"older exponents and to the finite size $N=5000$ of the sample. The width $\sigma$ of the histogram is expected to converge to zero as  $N \to \infty $.  

The problem is  to distinguish  fractal noise from  multifractal noise. The idea is that a multifractal noise is characterized by a   probability distribution of H\"older exponents {\it wider} than that of a correspondent monofractal noise. Therefore, with the help of Eq. (\ref{gaussf}) we suggest the following procedure for studying the multifractality of a time series of finite length $N$:
(i) we evaluate the mean H\"older exponent $\overline{h}$ and the width $\sigma$ of the histogram that estimates the probability distribution of the H\"older exponents of such datasets by using the Gaussian (\ref{gaussf});
(ii)  we generate many artificial datasets of fractal noise of finite length $N$ and with a Hurst coefficient $H=\overline{h}+1$ and study the  
 distribution of the monofractal widths $\sigma_F$; 
(iii) finally, if $\sigma$ is  larger than $\sigma_F$  and this is statistically significant, we  conclude that the original time series is multifractal.

%%%%%
\section{Human gait analysis}
We  now present the analysis of  the human gait of 10 persons in the three different conditions of slow, normal and fast walking  for a period of approximately one hour (unconstrained walking) and 30 minutes (metronomically constrained walking) in each condition. Figures 1 and 7 show two typical sets of data for  particular walkers under the three conditions.  

Participants in the study had no history of any neuromuscular, respiratory or cardiovascular disorders. They   were not taking any  medications and had  a mean age of 21.7 years (range: 18-29 years);  height $1.77 \pm 0.08$ meters  and mean weight  $71.8 \pm 10.7$ kg. All subjects provided informed written consent. 
Subjects walked continuously on level ground around an obstacle-free, long (either 225 or 400 meters), approximately oval path and the stride interval was measured using ultra-thin, force sensitive switches taped inside one shoe. For the metronomically constrained walking, the individuals were told only once, at the beginning of their walk, to synchronize their steps with the metronome.  More details regarding the collection of data can be found in  Physionet \cite{physionet} from where the data were downloaded and in Ref. \cite{Hausdorff4}. 

\subsection{Free-pace walking}

Table 1 records the basic properties of the 30 gait datasets at the unconstrained walking condition.  We tabulate the number of strides ($N$),  mean stride interval ($T$) and standard deviation of the stride interval for the three gait conditions for each of the ten walkers. The data condensed in Table 1 show a large variation in parameter values from  person to person. The mean value of the stride interval in the case of slow gait is $T=1.324$ sec., in the case of normal gait is $T=1.117$ sec., and in the case of fast gait is $T=1.001$ sec.. 
Person number 8 has the slowest walk with $T=1.790$ sec. and the standard deviation is the highest with $St~Dev=0.15$ sec.  Persons 1, 2 and 3 do not present large differences between slow and normal gait if we  focus on their mean stride interval time.

Table 2 shows the mean H\"older exponent $\overline{h}$ determined using  Eq. (\ref{linfitesa}) for the 30 gait datasets.   The fit is done on the scale  interval $1\leq s\leq 20$ that allows us to  explore  windows up to approximately 200 strides. 
Table 3 records the basic properties of the H\"older exponent distribution, explained in the previous section, for the 30 gait datasets. The mean  $h_0$ and the width of the distribution $\sigma$ are estimated by fitting  the histogram with the normalized Gaussian of Eq. (\ref{gaussf}) as done in Fig. 6 for  computer generated FGN. The error of measure on the mean value $h_0$  is estimated  to be  $\pm 0.0025$ on average, whereas the error on the width $\sigma$ is  $\pm 0.002$ on average. 

Table 3 also records the width $\sigma_F$  of the distribution of H\"older exponents for computer-generated  datasets of monofractal noise with $H=1+\overline{h}$ of $N$ elements in correspondence of  each of the 30 gait datasets. The values $\sigma_F$ are evaluated by averaging 20 computer simulations.
The table shows that, even by considering the error of measure of $\pm0.002$,  the width of the H\"older exponent distribution $\sigma$ is almost always slightly larger than  the width $\sigma_F$ for a corresponding monofractal noise of Hurst exponent $H=\overline{h}+1$  of equal size sample $N$. On average,  we get that for slow gait $\sigma$ is $8.3\%$ larger than $\sigma_F$, for the normal gait $\sigma$ is   $4.7\%$ larger than $\sigma_F$ and, finally, for the fast gait $\sigma$ is $4.9\%$  larger than the correspondent $\sigma_F$. In particular, for the slow gait of  person number 8, see Figs. 7 and 8B,   $\sigma$ is  $40\%$ larger than $\sigma_F$.  

Table 4 reports the results of the Student's t-test in the case of paired samples \cite{nr} between the histogram widths $\sigma$ and $\sigma_F$ for the 10 walkers in the three gait conditions  recorded in Table 3. 
We use the paired samples algorithm because the values of the widths $\sigma$ and $\sigma_F$ depend on the size of the sample $N$ and the Hurst exponent $H$.  So, we have to imagine that the variance in both samples may be due to effects that are point-by-point identical in the two samples.
The value $t$ is the Student's t value and $prob$  is the probability that the two sets of data for each walking condition have the same mean.
Figs. 8A and 8B show the probability density function of the H\"older exponents for  walker number 5 (Fig. 8A) and  walker number 8 (Fig. 8B). Finally, Fig. 9 shows the global distribution of the H\"older exponents for the three different gaits whose characteristics are condensed in Table 5. Three symbols --star, triangle and circle-- indicate the three gaits --slow, normal and fast--.  The distributions are fitted by normalized Gaussian functions, Eq. (\ref{gaussf}).

Fig. 9 and Table 5 show the global properties of the distributions of all H\"older exponents for the three different gait modes  that have been measured for the ten persons.  By increasing the average rate of walking from slow to normal the mean of the Gaussian, $h_0$,  on average decreases, whereas increasing the average rate of walking from normal to fast, $h_0$ increases on  average.  There is also an increasing of the width of the distribution $\sigma$ by moving from the normal to the slow or fast gait mode. This last result indicates that  normal human  gait is more standard than the other two types of gait in the sense that many persons, when asked to walk normally, present a similar distribution of  H\"older exponents for the stride interval time series. Moreover, we note a large width of the distribution of the H\"older exponents in the case of slow gait. This means that there is  a large variability in the distribution of H\"older exponents  for slow human gait, that is, a large variability of the fractal properties of the stride interval time series among the persons who are asked to walk slowly.

\subsection{Metronomically-pace walking}

Figs. 10 show two of the ten individuals of metronomically constrained walking for the three gait conditions; slow, normal and fast walking.
The total period for each dataset is  approximately 30 minutes. Table 6 records the basic properties of the 30 gait datasets. The mean values of the stride interval is compatible with those obtained for the unconstrained walking: in the case of slow gait  $T=1.356$ sec.; in the case of normal gait  $T=1.116$ sec.; and in the case of fast gait  $T=1.003$ sec.. However, by comparing Tables 1 and 6 as well as  Figs. 1, 7 and 10, we notice that the constrained walking presents a smaller standard deviation and  much less variability of the strength of the local biases of the stride interval time series. This can be  understood as an effect due to the unvaryingly regular artificial rhythm that constrains the walking.

Table 7 shows the mean H\"older exponent $\overline{h}$ determined using  Eq. (\ref{linfitesa}) for the 30 gait datasets.   The fit is done on the scale  interval $1\leq s\leq 10$ that allows us to  explore  windows up to approximately 100 strides.  We use a scale interval up to $s=10$ because the fewer number of data points (almost one half of the previous case) makes  the statistics poorer at higher scales and because, here,  we  study the differences that occurs among the three gait conditions at the shorter scale. Figs. 11A and 11B show the histograms of the H\"older exponents for  walker number 3 (Fig. 11A), who can be considered  to be typical of the ten walkers, and  walker number 5 (Fig. 11B) characterized by a strong antipersistent behavior of the stride interval time series. Finally, Fig. 12 shows the global distribution of the H\"older exponents as  results from the study of the 30 datasets for the three different gait modes.  The figure shows  wide spreading of the global distribution of H\"older exponents, a fact that suggests a large variability of situations from persistent to antipersistent conditions.  

Table 8 reports the  values $h_0$ and $\sigma$ of the  normalized Gaussian functions, Eq. (\ref{gaussf}),  that fit the histograms of the H\"older exponent probability density distribution. The values $\sigma_F$ are the estimated width of the H\"older exponent probability density distribution of the computer-generated monofractal noise with Hurst exponent $H=1+\overline{h}$ and of $N$ elements. Finally, Table 9 records the Student's t-test is the case of paired samples between the histogram widths $\sigma$ and $\sigma_F$ for the 10 walkers in the three gait conditions. The value $t$ is the Student's t value and $prob$  is the probability that the two sets of data for each walking condition have the same mean. Table 8 and the Student's t-test shows that the widths $\sigma$ are  usually larger than the correspondent $\sigma_F$ and this increase is statistically significant.

%%%%%
\section{Discussion and conclusion}
Hausdorff et al. \cite{Hausdorff4} established that during the metronomically-paced walking, the long-range correlations of up to 1000 strides detected in the three modes of free walking disappear and variations in the stride interval are anti-correlated. These results are confirmed by  the present analysis. However, the study of the distribution of the H\"older exponents allows for an even richer interpretation of the scaling behavior of the inter-stride interval time series. The time series is not monofractal, as was suggested by earlier analysis \cite{hausdorff1,griffin}, but is here determined to be weakly multifractal. The multifractality does not strictly invalidate the interpretation of scaling behavior, that being, that the statistical correlations in the stride interval fluctuations over thousands of strides decay in a scale-invariant manner. But it does suggest that the scale-invariant decay of the correlations is more complicated than was previously believed.

The average H\"older exponent, or equivalently, the fractal dimension, is determined to be dependent on the average rate at which an individual walks, but not monotonically dependent. The fractal dimension for the fast gait lies between those for the slow and normal gaits, in the case of unconstrained walking. The ordering of the fractal dimension for the three modes of walking is not so evident for the metronomically constrained gait.

We note that in the case of unconstrained walking the standard deviation in the case of slow gait is usually larger, almost double,  that of the  fast and slow gait cases.  One possible explanation of this larger variance is that  the stride interval for slow gait may be characterized  by a non-stationary change  in the mean stride interval during walking, what Hausdorff et al. [4] refer to as loss of concentration. This non-stationarity is seen to be the case in Fig. 1 and more specifically in Fig. 7, for one individual, but this behavior is typical of all the walkers. It is also worth pointing out that the standard deviation of the stride interval fluctuations for the slow gait increases as the mean stride interval decreases. This suggests that the more slowly a person walks, the  more difficult it is for that person to keep his/her gait regular. 

 The results condensed in Tables 2 and 3, and in Figs. 8 and 9 show   a great deal of information about the fractal properties of human gait. Note that the mean H\"older exponent $\overline{h}$ is usually slightly smaller than the center of the Gaussian fitting distribution $h_0$ because the H\"older exponent distributions present a slight positive skewness.   All distributions of H\"older exponents are  centered very close to the value $h=0$ that characterizes  pink noise. Normal gait always presents  a negative mean  H\"older exponent $\overline{h}$ but larger than $h=-0.2$. This fact indicates that the stride interval time series for normal gait is  strongly persistent and stationary, characterized by long-range, fractal correlations.  

Fast gait   presents properties similar to those of the normal gait,  but usually with a  mean  H\"older exponent slightly larger and  closer to the threshold $h=0$. The fact that fast gait  almost always presents a negative mean  H\"older exponent $\overline{h}$ means that the stride interval time for fast gait can usually be considered to be strongly persistent noise that,  and as in the previous case,  is characterized by strong  long-range, fractal correlations.  By contrast, at least  for two people (persons number 1 and 2), the threshold $h=0$ of the pink noise is surpassed. This means that in these two cases the stride interval time fluctuations are slightly non-stationary. This last result emphasizes the range of dynamics of normal healthy gait.

On the contrary, the stride interval time series for slow gait,   usually presents a mean  H\"older exponent $\overline{h}$ slightly larger than the pink noise threshold $h=0$. This  shift in the peak of the distribution to more positive Holder exponents implies that  the slow gait is usually characterized by non-stationary fluctuations of the stride interval time series. Therefore, the slow gait regime could be considered  a strongly anti-persistent and non-stationary {\it walk}  rather than  a strongly persistent {\it noise}.  Perhaps, this slight non-stationarity in the slow gait is related to the fact that, contrary to the normal and, in part, the fast gait, walking slowly may require more concentration and a person asked to walk slowly  may unconsciously lose this concentration and change the way of walking as he or she feels more comfortable.  

The comparison between the probability density widths $\sigma$ for the gait data and $\sigma_F$ for the correspondent monofractal noise, that are recorded in the  Tables 3 and 4, supports the conclusion that human gait may be characterized on average by a form of multifractality. In fact, the slowest mode of walking has the most variability as measured by the  relative width $\sigma/\sigma_F$. The fastest mode of walking  is characterized by a relative width compatible to or slightly larger than that of the normal mode.

The metronomically constrained walking datasets presents  more complex behavior than does the freely walking data. The values of the mean H\"older exponents $\overline{h}$ recorded in Table 7 and Fig. 12 show that the walking loses the strong persistence of the unconstrained gait and becomes more random ($\overline{h}\approx -0.5$) or antipersistent ($\overline{h}< -0.5$). However, the normal gait still presents  persistent behavior for many individuals. This indicates that   spontaneous walking is less influenced by external constraints than  either the fast of slow walking conditions. We stress the fact that for constrained gait our analysis concerns windows of  width up to 100 strides.  The fast gait becomes more random on average. This may indicate that a synchronization of the walking is easier in the fast regime.  The slow gait shows a wider spectrum of situations from persistent to antipersistent noise, $-0.9<\overline{h}<-0.1$  indicating that synchronization of the walking is more difficult for some people in the slow regime. 

We notice that at least  one person, walker number 5, presents a strong antipersistency, $\overline{h}<-0.9$, for each of the three gait conditions.   This may indicate that this person,  in trying to synchronize his walking to the frequency of the clock,   is unable to find a standard condition at all three gait  modes.  Consequently,  the walker continuously shifts his stride interval up and down in the vicinity of an average, giving rise to a strong antipersistent signal, see also Fig. 11B. Paradoxically, the antipersistence of the signal is strongest at the normal gait. This effect may be a consequence of  normal gait  being free from the supraspinal influences of the metronome. This individual finds it necessary to continuously readjust his walking to maintain synchrony with the metronome.

It is well known to everyone that has taken military basic training that there are some individuals who cannot march in cadence. Individual 5 seems to suffer from this particular malady, but this interpretation requires additional research. 

Finally, Tables 8 and 9 allow the comparison between the probability density widths $\sigma$ for the gait data and $\sigma_F$ for the correspondent monofractal noise. The width $\sigma$ are often larger than the correspondent $\sigma_F$ for all three conditions and the Student's t-test allows us to conclude that this difference is statistically significant.  This supports the conclusion that  the human gait  is characterized on average by a form of multifractality  for the metronomically constrained walking. Also in this case, the fast and the slow gait are likely to be  more multifractal than normal gait.  

We noted earlier that there are a number of mathematical versions of the Central Patten Generators (CPGs) used to model the groups of neurons producing the rhythmic signals that produce locomotion in animals and possibly in humans as well. Here we note that certain coupled nonlinear oscillator networks have trajectories that lie on strange attractors. In one such case the time series resulting from such solutions have been shown to be multifractal, which is to say, a singularity spectrum was calculated from the time series \cite{nakamura}. The properties of the solution to such a nonlinear dynamical system appears to be consistent with the processing results obtained herein for gait. For example, Nakamura \cite{nakamura} showed that the singularity spectrum has multiple scaling regions (peaks in the singularity spectrum) dependent on certain parameter values in the dynamical equations. Of course it is necessary to provide a physiological interpretation of the parameters in this nonlinear oscillator before making any claims as to applicability as a CPG model. We are presently exploring that possibility.

--------\\
{ {\large \bf Acknowledgment:}}\\
N.S. thanks the Army Research Office for support under grant DAAG5598D0002 and L.G. thanks the National Research Center Fellowship.

%%%%

%%%%%%%%%%%

\onecolumn

%%%%%%%%%%%

\newpage
\begin{table}
  \begin{tabular}{|c|ccc|ccc|ccc|}\hline
          &      & Slow&      &      &Norm&      &      &Fast&      \\ 
 walker&N    &$T$&$St~ Dev$&N&$T$&$St~ Dev$&N&$T$&$St~ Dev$      \\ \hline
 1         &3304      &1.167      &0.03      &3371      &1.037      &0.02      &3595      &1.006      &0.02      \\ \hline
 2         &3347      &1.063      &0.02      &3357      &0.964      &0.02      &3822      &0.925      &0.02       \\ \hline
3          &3257      &1.088      &0.02      &3391      &1.078      &0.02      &3517      &0.979      &0.01     \\ \hline
 4         &2625      &1.372      &0.05      &3126      &1.124      &0.02      &3534      &1.008      &0.02      \\ \hline
 5         &2496      &1.461      &0.05      &3362      &1.106      &0.02      &3819      &0.948      &0.03      \\ \hline
  6        &2844      &1.273      &0.04      &3297      &1.108      &0.02      &3451      &1.008      &0.02      \\ \hline
7          &2717      &1.338      &0.06      &2976      &1.149      &0.02      &3447      &1.058      &0.02      \\ \hline
 8         &2040      &1.790      &0.15      &2902      &1.183      &0.02      &3720      &0.967      &0.01      \\ \hline
 9         &2764      &1.315      &0.02      &3054      &1.179      &0.02      &3447      &1.042      &0.01      \\ \hline
10        &2547      &1.373      &0.04      &2977      &1.242      &0.02      &3262      &1.070      &0.02      \\ \hline 
Ave.     &2794   &  1.324       &0.05      &3181      & 1.117     &0.02      &3561      &1.001      &0.02      \\ \hline
  \end{tabular} 
\caption{Number of strides (N),  mean stride interval (T) and standard deviation of the stride interval for the ten walkers. Each person walks for approximately  one hour in each of three different modes: slow, normal and fast. }
\end{table}

\begin{table}
  \begin{tabular}{|c|c|c|c|}\hline
          &   Slow      &    Norm      &    Fast      \\ 
 walker&   $\overline{h} $	&$\overline{h}	$	&$\overline{h} $ 	   \\ \hline
 1         &$ -0.013 \pm 0.011$      &$ -0.068\pm 0.015$      &$0.030 \pm 0.007$          \\ \hline
 2         &$ -0.163 \pm 0.008 $      &$-0.114 \pm 0.012$     &$ 0.027 \pm 0.021  $            \\ \hline
3          &$ 0.030\pm 0.018$      &$ -0.093 \pm 0.011$      &$ -0.099\pm 0.018$           \\ \hline
 4         &$ 0.087\pm 0.016$      &$ -0.117\pm0.014 $      &$ -0.077\pm0.012 $            \\ \hline
 5         &$ 0.101\pm0.013 $      &$ -0.132\pm 0.012$      &$ -0.020\pm0.009 $            \\ \hline
 6        &$ 0.090\pm 0.017$      &$ -0.081\pm0.018 $      &$ -0.134\pm0.011 $            \\ \hline
7          &$ 0.172\pm0.032 $      &$ -0.096\pm0.016 $      &$ -0.044\pm0.010 $            \\ \hline
 8         &$ 0.034\pm0.017 $      &$ 0.002\pm0.015 $      &$ -0.031\pm 0.019$            \\ \hline
 9         &$ 0.003\pm0.017 $      &$ -0.085\pm0.017 $      &$ -0.114\pm0.011 $            \\ \hline
10        &$ -0.056\pm 0.014$      &$ -0.188\pm 0.008$      &$ -0.027\pm0.010 $            \\ \hline 
Ave.     &$ 0.028\pm 0.089$  &  $ -0.097\pm 0.049$       &$ -0.049\pm0.056 $            \\ \hline
  \end{tabular} 
\caption{Mean H\"older exponent $\overline{h}$ given by Eq. (\ref{linfitesa}) for the 30 gait datasets.   The fit is done on the scale  interval $1\leq s\leq 20$. The exponent $\overline{h}$ is related to the Hurst exponent $H$ via the relation $\overline{h}=H-1$.   }
\end{table}

\begin{table}
  \begin{tabular}{|c|c c c|c c c|c c c|}\hline
          &      &Slow & & &Norm      & & &Fast     &       \\ 
walker & $h_0$  & $\sigma$ &$\sigma_F$& $h_0$ & $\sigma$ &$\sigma_F$& $h_0$ & $\sigma$&$\sigma_F$      \\ \hline
1         &-0.009    &0.059	 & 0.056   &-0.063      &0.062	& 0.055     &0.041      &0.059& 0.055    \\ \hline
2         &-0.154    &0.057	&  0.056   &-0.110      &0.053	& 0.055     &0.019      &0.060& 0.054     \\ \hline
3         &0.026     &0.059 	&  0.056   &-0.090      &0.056	& 0.055     &-0.095     &0.061& 0.055     \\ \hline
4         &0.090     &0.058 	& 0.056   &-0.105       &0.056	& 0.056     &-0.072     &0.054& 0.055     \\ \hline
5         &0.105     &0.060 	& 0.057    &-0.125      &0.063	&  0.055    &-0.012     &0.056&  0.054    \\ \hline
6         &0.088     &0.060	&  0.056    &-0.083     &0.059	& 0.056     &-0.128     &0.059&  0.055    \\ \hline
7         &0.161     &0.061	&  0.056    &-0.089     &0.060	&  0.056    &-0.035     &0.057&  0.055    \\ \hline
8         &0.075     &0.081	&   0.058   &-0.000     &0.058	&  0.056    &-0.040     &0.055&  0.054    \\ \hline
9         &-0.002    &0.052	&  0.056   &-0.090      &0.057	& 0.056     &-0.114     &0.054& 0.055     \\ \hline
10       &-0.031    &0.064	&  0.057   &-0.165      &0.058	&  0.056    &-0.011     &0.060& 0.056     \\ \hline
 Ave.   & 0.035    &0.0611  	&  0.0564   & -0.092   &0.0582      	& 0.0556   &-0.045     &0.0575 &0.0548 	\\ \hline
  \end{tabular} 
\caption{H\"older exponents distribution for the ten walkers. The distributions are fitted by Gaussian functions where $h_0$ is the mean and $\sigma$ is the standard deviation. The error of measure on the mean value $h_0$  is estimated in average to be $\pm 0.003$, whereas the error on the widths $\sigma$ and $\sigma_F$ is on average $\pm 0.002$. The width $\sigma_F$ is the estimated width of the distribution of H\"older exponents for a dataset of computer-generated monofractal noise with $H=1+\overline{h}$ of $N$ elements.  }
\end{table}

\begin{table}
  \begin{tabular}{|c|c|c|c|}\hline
  t-test     & slow  & normal & fast       \\ \hline
$t$    & 2.11     &  2.68   & 3.36        \\ \hline
$prob$     & 0.064     &   0.025   &  0.008       \\ \hline
 \end{tabular} 
\caption{ Student's t-test in the case of paired samples between the histogram widths $\sigma$ and $\sigma_F$ for the 10 walkers in the three gait conditions, see Table 3. The value $t$ is the Student's t value and $prob$  is the probability that the two sets of data for each walking condition have the same mean.   }
\end{table}

\begin{table}
  \begin{tabular}{|c|c|c|}\hline
gait        & $h_0$  & $\sigma$       \\ \hline
slow         &$0.046 \pm 0.002$      & $0.102 \pm 0.001$        \\ \hline
norm         &$-0.092 \pm 0.001$      &$0.069 \pm 0.001$       \\ \hline
fast         &$-0.035 \pm 0.001$      &$0.081 \pm 0.001$         \\ \hline
  \end{tabular} 
\caption{The mean value $h_0$ and the standard deviation $\sigma$ of H\"older exponent distribution of the ten walkers in the three speed gait cases. }
\end{table}

%%%%%%

\begin{table}
  \begin{tabular}{|c|ccc|ccc|ccc|}\hline
          &      & Slow&      &      &Norm&      &      &Fast&      \\ 
 walker&N    &$T$&$St~ Dev$&N&$T$&$St~ Dev$&N&$T$&$St~ Dev$      \\ \hline
 1         &1508      &1.167      &0.02      &1651      &1.046      &0.01      &1804      &1.010      &0.01      \\ \hline
 2         &1705      &1.061      &0.02      &1797      &0.961      &0.01      &1781      & 0.932     &0.01       \\ \hline
3          &1357      &1.336      &0.03      &1652      &1.083      &0.01      &1900      & 0.986     &0.02     \\ \hline
 4         &1416      &1.367      &0.03      &1703      &1.124      &0.02      &1839      &1.013      &0.02      \\ \hline
 5         &1306      &1.465      &0.03      &1586      &1.114      &0.01      &1956      &0.954      &0.01      \\ \hline
 6         &1430      &1.279      &0.03      &1638      &1.113      &0.01      &1765      &1.010      &0.01       \\ \hline
7          &1415      &1.302      &0.04      &1734      &1.113      &0.01      &1723      &1.046      &0.01      \\ \hline
 8         &1210      &1.795      &0.04      &1438      &1.190      &0.02      &1791      &0.962      &0.01      \\ \hline
 9         &1410      &1.321      &0.02      &1573      &1.178      &0.02      &1683      &1.045      &0.01      \\ \hline
10        &1390      &1.365      &0.02      &1530      &1.239      &0.02      &1596      &1.073      &0.01      \\ \hline 
Ave.     &1415      &1.356      &0.03      &1630      &1.116      &0.02      &1784      &1.003      &0.01      \\ \hline
  \end{tabular} 
\caption{Metronomic walking. Number of strides (N),  mean stride interval (T) and standard deviation of the stride interval for the ten walkers. Each person walks for approximately  one-half hour in three different ways: 1) slow; 2) normal; 3) fast. }
\end{table}

\begin{table}
  \begin{tabular}{|c|c|c|c|}\hline
          &   Slow      &    Norm      &    Fast      \\ 
 walker&   $\overline{h} $	&$\overline{h}	$	&$\overline{h} $ 	   \\ \hline
 1         &$ -0.479 \pm0.039  $      &$-0.248  \pm 0.025  $      &$-0.405  \pm0.031  $          \\ \hline
 2         &$-0.167  \pm0.006  $      &$-0.541  \pm 0.040 $      &$-0.470  \pm0.023  $            \\ \hline
3          &$-0.827  \pm0.060  $      &$-0.245  \pm 0.040 $      &$-0.505  \pm0.052  $           \\ \hline
 4         &$-0.769  \pm 0.087  $      &$-0.599  \pm0.077  $      &$-0.118  \pm0.024  $            \\ \hline
 5         &$-0.931  \pm 0.037 $      &$-1.121  \pm 0.107  $      &$-0.945  \pm 0.042  $            \\ \hline
 6         &$-0.552  \pm0.060  $      &$-0.432  \pm0.038  $      &$-0.459  \pm 0.012  $            \\ \hline
7          &$-0.277  \pm0.014  $      &$-0.196  \pm0.010  $      &$-0.370  \pm 0.014  $            \\ \hline
 8         &$-0.894  \pm0.080  $      &$-0.347  \pm 0.037  $      &$-0.151  \pm0.035  $            \\ \hline
 9         &$-0.292  \pm0.053  $      &$-0.176  \pm 0.019  $      &$-0.228  \pm0.020  $            \\ \hline
10        &$-0.179  \pm0.027  $      &$-0.183  \pm 0.022  $      &$-0.469  \pm0.022  $            \\ \hline 
Ave.     &$-0.537  \pm 0.301 $      &$-0.409  \pm 0.292 $       &$-0.412  \pm 0.234 $            \\ \hline
  \end{tabular} 
\caption{Metronomic walking. Mean H\"older exponent $\overline{h}$ given by Eq. (\ref{linfitesa}) for the 30 gait datasets.   The fit is done over the scale  interval $1\leq s\leq 10$. The exponent $\overline{h}$ is related to the Hurst exponent $H$ via $\overline{h}=H-1$.   }
\end{table}

\begin{table}
  \begin{tabular}{|c|c c c|c c c|c c c|}\hline
          &      &Slow & & &Norm      & & &Fast     &       \\ 
walker & $h_0$  & $\sigma$ &$\sigma_F$& $h_0$ & $\sigma$ &$\sigma_F$& $h_0$ & $\sigma$&$\sigma_F$      \\ \hline
1         &-0.439     &0.063	&0.061    &-0.224     &0.062	&0.060    &-0.366      &0.067   &0.058      \\ \hline
2         &-0.153     &0.066	&0.061    &-0.510     &0.061	&0.059    &-0.444      &0.057   &0.059      \\ \hline
3         &-0.765     &0.064 	&0.063    &-0.204     &0.064	&0.060    &-0.436      &0.066   &0.059      \\ \hline
4         &-0.712     &0.058 	&0.062    &-0.542     &0.069	&0.060    &-0.058      &0.072   &0.058      \\ \hline
5         &-0.895     &0.066 	&0.064    &-1.058     &0.059	&0.063    &-0.902      &0.066   &0.059      \\ \hline
6         &-0.492     &0.071	&0.060    &-0.392     &0.060	&0.060    &-0.429      &0.067   &0.060      \\ \hline
7         &-0.164     &0.071	&0.062    &-0.174     &0.059	&0.060    &-0.250      &0.065   &0.059      \\ \hline
8         &-0.818     &0.069	&0.064    &-0.272     &0.067	&0.061    &-0.131      &0.060   &0.059      \\ \hline
9         &-0.251     &0.062	&0.062    &-0.143     &0.069	&0.060    &-0.203      &0.058   &0.060      \\ \hline
10       &-0.148     &0.068	&0.061    &-0.154     &0.062             &0.061    &-0.425      &0.072   &0.059      \\ \hline
 Ave.   &-0.484     &0.066  	&0.0622    &-0.367   &0.063      	&0.0604  &-0.364      &0.065   &0.0590 	     \\ \hline
  \end{tabular} 
\caption{Metronomic walking. H\"older exponents distribution for the ten walkers. The distributions are fitted by Gaussian functions where $h_0$ is the mean and $\sigma$ is the standard deviation. The error of measure on the mean value $h_0$  is estimated  to be $\pm 0.005$ on average, the error on the widths $\sigma$ and $\sigma_F$ is  $\pm 0.003$ on average. The width $\sigma_F$  is the estimated width of the distribution of H\"older exponents for a dataset of monofractal noise with $H=1+\overline{h}$ of $N$ elements.  }
\end{table}

\begin{table}
  \begin{tabular}{|c|c|c|c|}\hline
  t-test     & slow  & normal & fast       \\ \hline
$t$          &2.68     &2.09    &3.41         \\ \hline
$prob$    &0.025     &0.066    &0.008        \\ \hline
 \end{tabular} 
\caption{ Metronomic walking. Student's t-test in the case of paired samples between the histogram widths $\sigma$ and $\sigma_F$ for the 10 walkers in the three gait conditions, see Table 3. The value $t$ is the Student's t value and $prob$  is the probability that the two sets of data for each walking condition have the same mean.   }
\end{table}

%%%%%%%%
\newpage

\begin{figure}

Figure 1\\
\epsfig{file=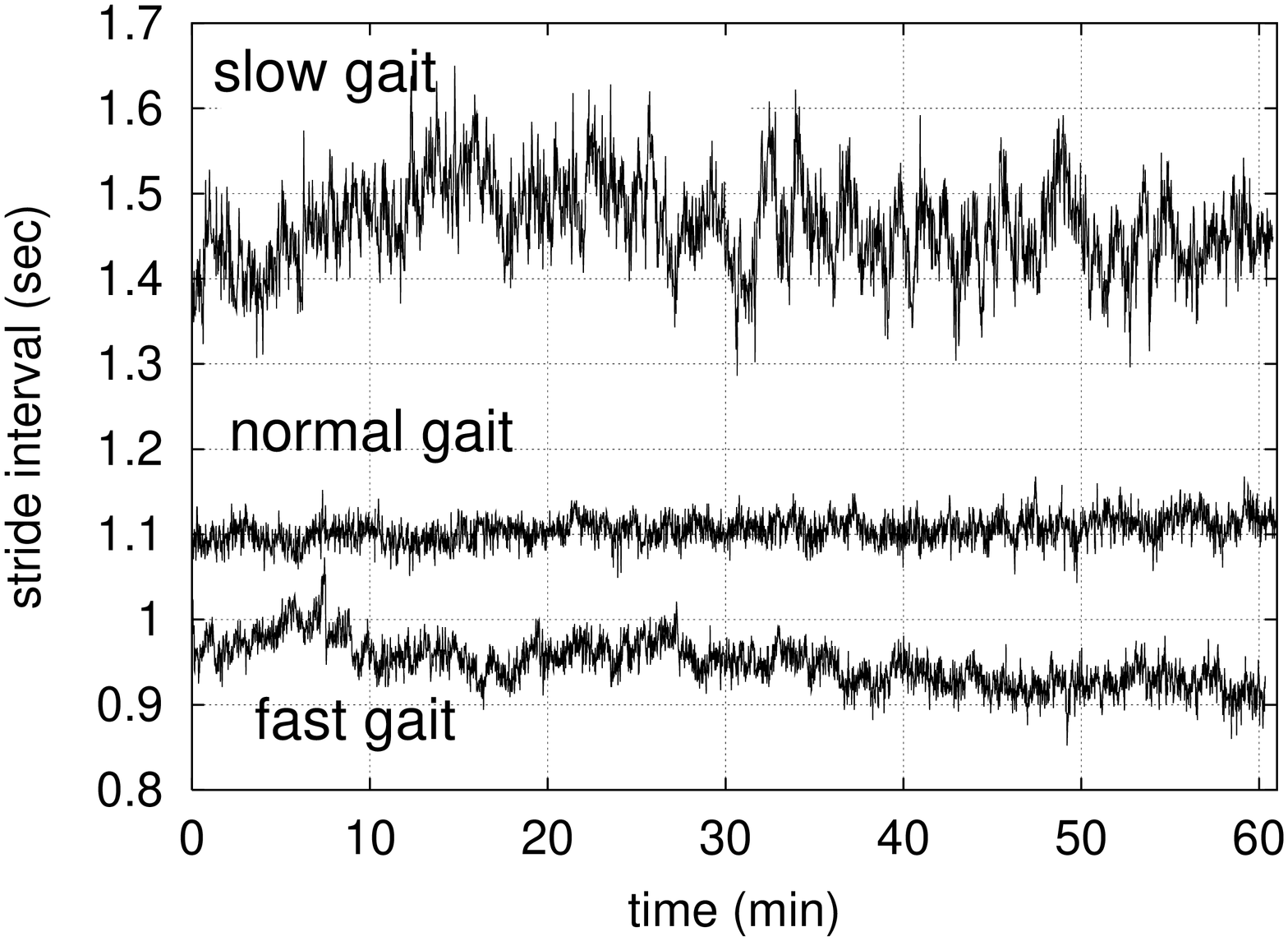,height=13cm,width=10cm,angle=-90}
\caption{ Stride interval for slow, normal and fast gait. The period of time over which  measurements were done is  approximately one hour. These are the data from person number 5.    }
\end{figure}

\newpage

\begin{figure}

Figure 2\\
\epsfig{file=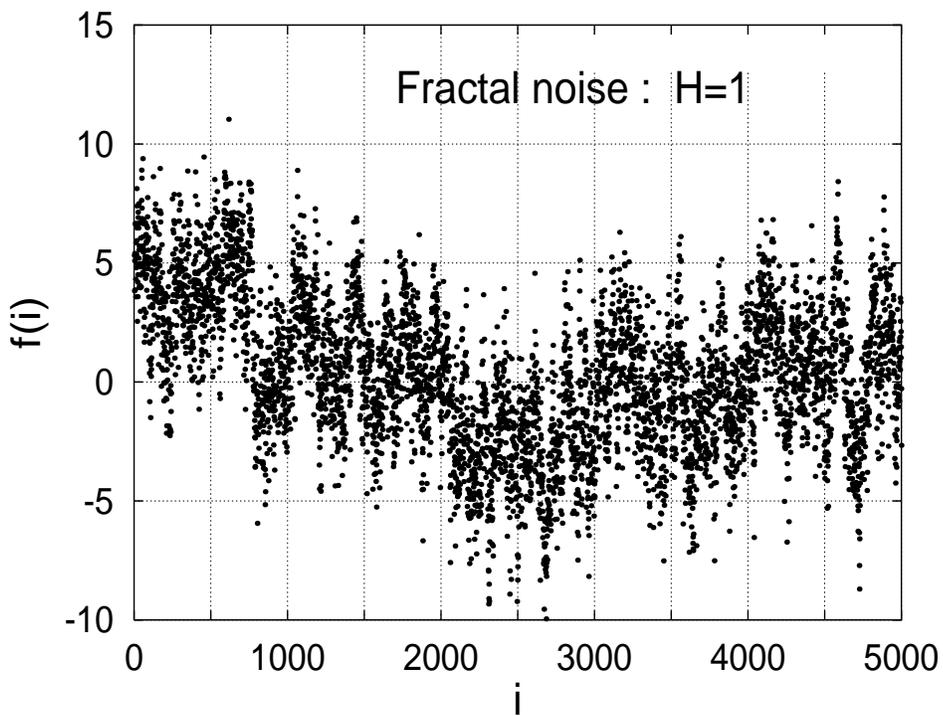,height=13cm,width=10cm,angle=-90}
\caption{ Computer-generated fractal noise (FGN) with Hurst coefficient $H=1$, also known as $1/f$ noise or {\it pink} noise is shown.    }
\end{figure}

\newpage

\begin{figure}

Figure 3\\
\epsfig{file=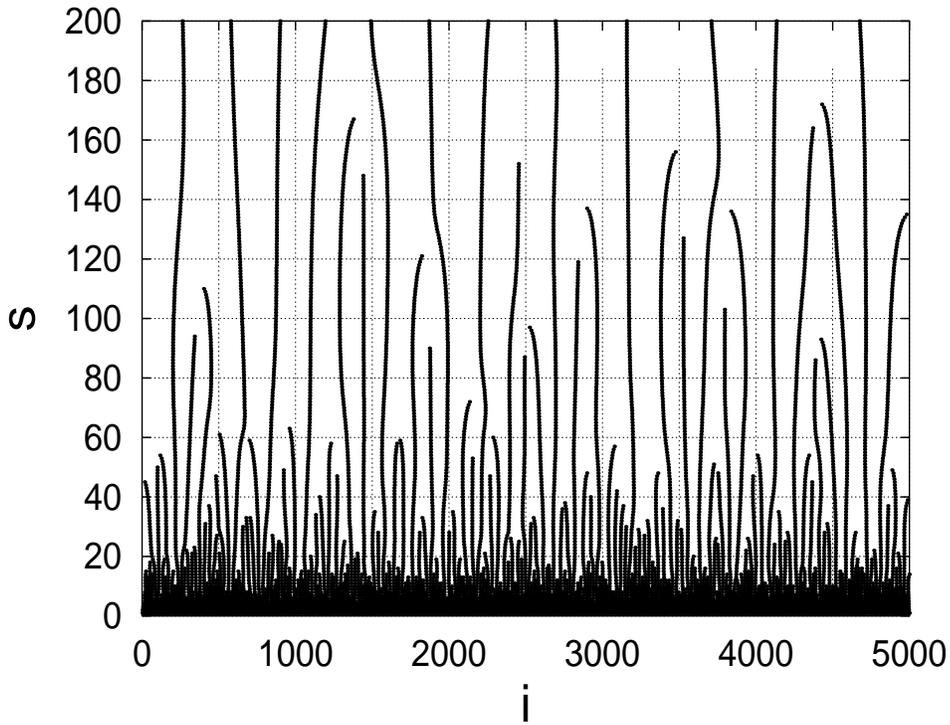,height=13cm,width=10cm,angle=-90}
\caption{ Wavelet transform modulus maxima lines of the computer-generated fractal noise of Fig. 2 are shown.    }
\end{figure}

\newpage

\begin{figure}

Figure 4\\
\epsfig{file=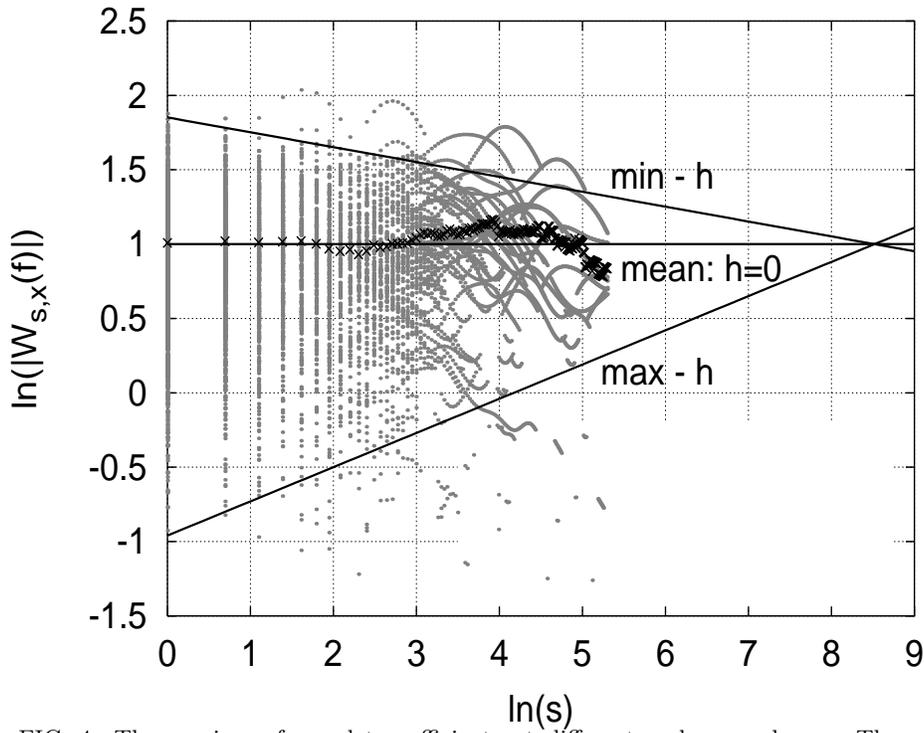,height=13cm,width=10cm,angle=-90}

\caption{The maxima of wavelet coefficients at different scales are shown. The coefficients are bounded in the interval $-1.2<h<1.2$,  [25]. The crosses indicate the mean value evaluated by using Eq. (\ref{linfitesa}). The intersection point at the right is at the maximum available scale $\log(5000)=8.52$. The three straight lines  indicate the slope of the mean, the maximum and minimum values of the H\"older exponents.}
\end{figure}

\newpage

\begin{figure}

Figure 5\\
\epsfig{file=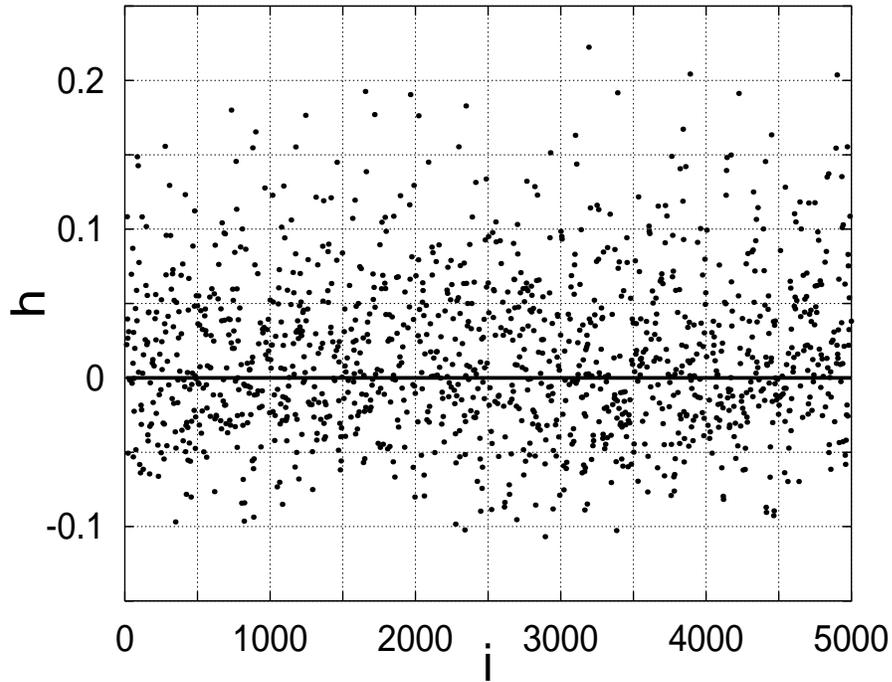,height=13cm,width=10cm,angle=-90}
\caption{  The H\"older exponents $h$ for the FGN are plotted against the position of the time series. The straight line correspond to the center of the distribution.    }
\end{figure}

\newpage

\begin{figure}

Figure 6\\
\epsfig{file=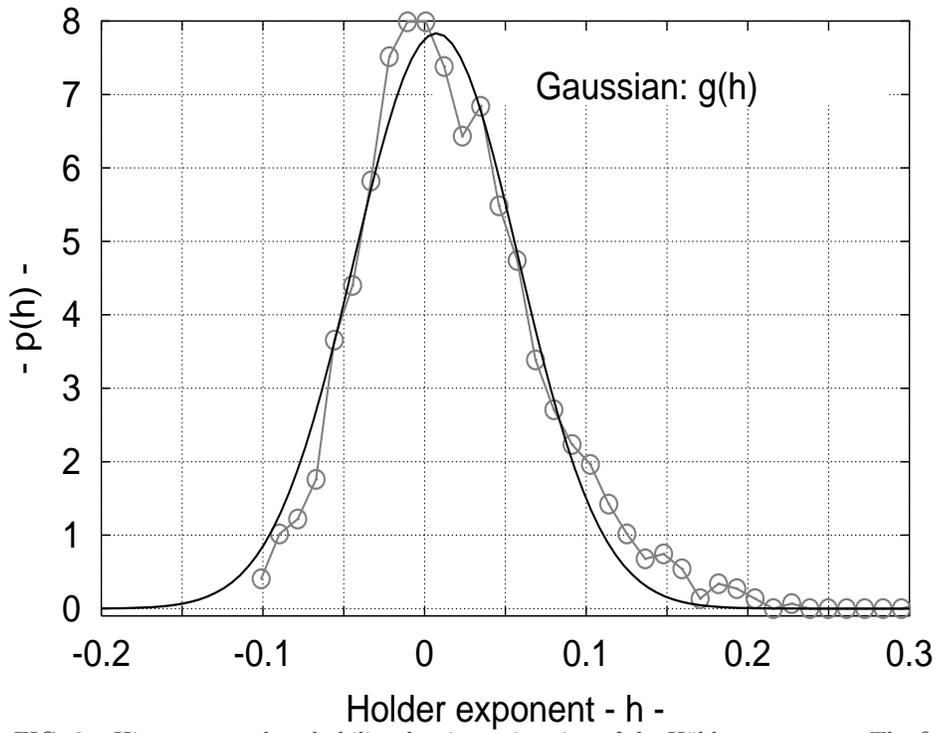,height=13cm,width=10cm,angle=-90}
\caption{  Histogram and probability density estimation of the H\"older exponents. The fitting curve is a Gaussian (\ref{gaussf})  centered in $h_0=0.007 \pm 0.001$ and with a width $\sigma=0.051 \pm 0.001$. The mean H\"older exponent given by Eq. (\ref{linfitesa})  is $\overline{h}=-0.004 \pm 0.008$ and correspond to the maximum of the distribution.      }
\end{figure}

\newpage

\begin{figure}

Figure 7\\
\epsfig{file=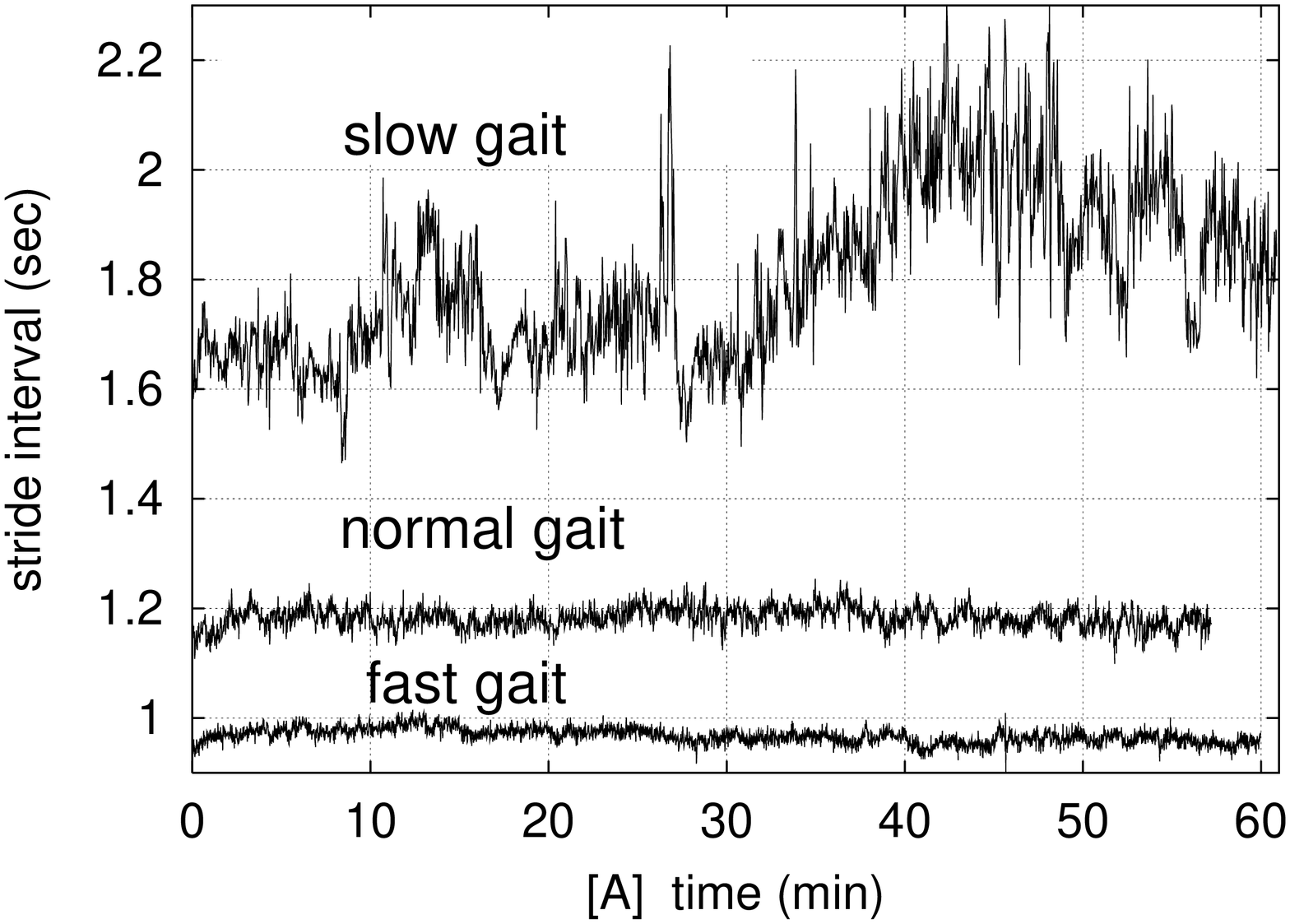,height=12cm,width=8cm,angle=-90}\\
\epsfig{file=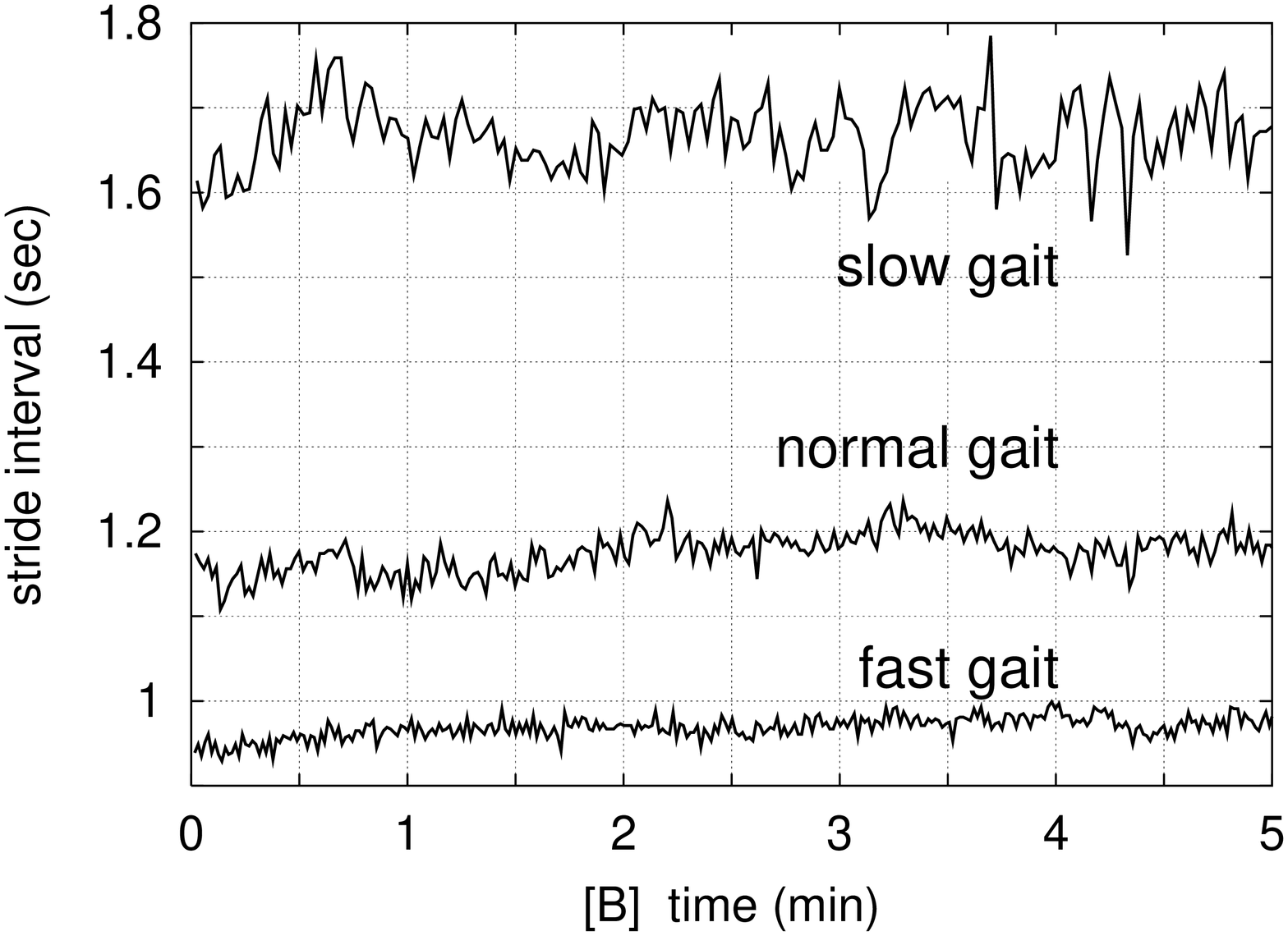,height=12cm,width=8cm,angle=-90}
\caption{ Stride interval for slow, normal and fast gait  are shown. (A) The total period of time is approximately 1 hour. (B) The total period of time is 5 minutes.  Note the variability of the fluctuation of the slow gait.  The data is for person number 8.    }
\end{figure}

\newpage

\begin{figure}

Figure 8\\
\epsfig{file=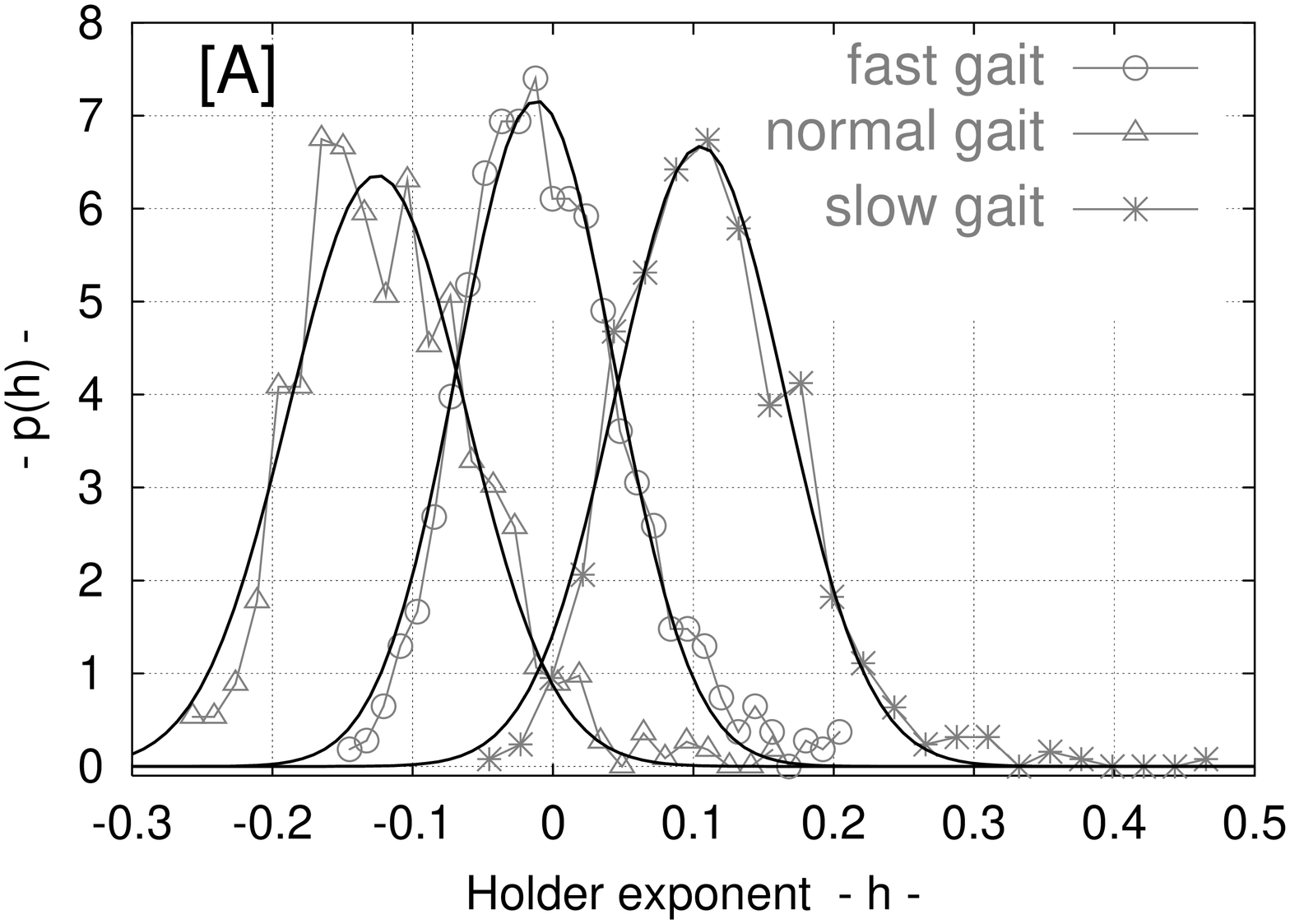,height=12cm,width=8cm,angle=-90}\\
\epsfig{file=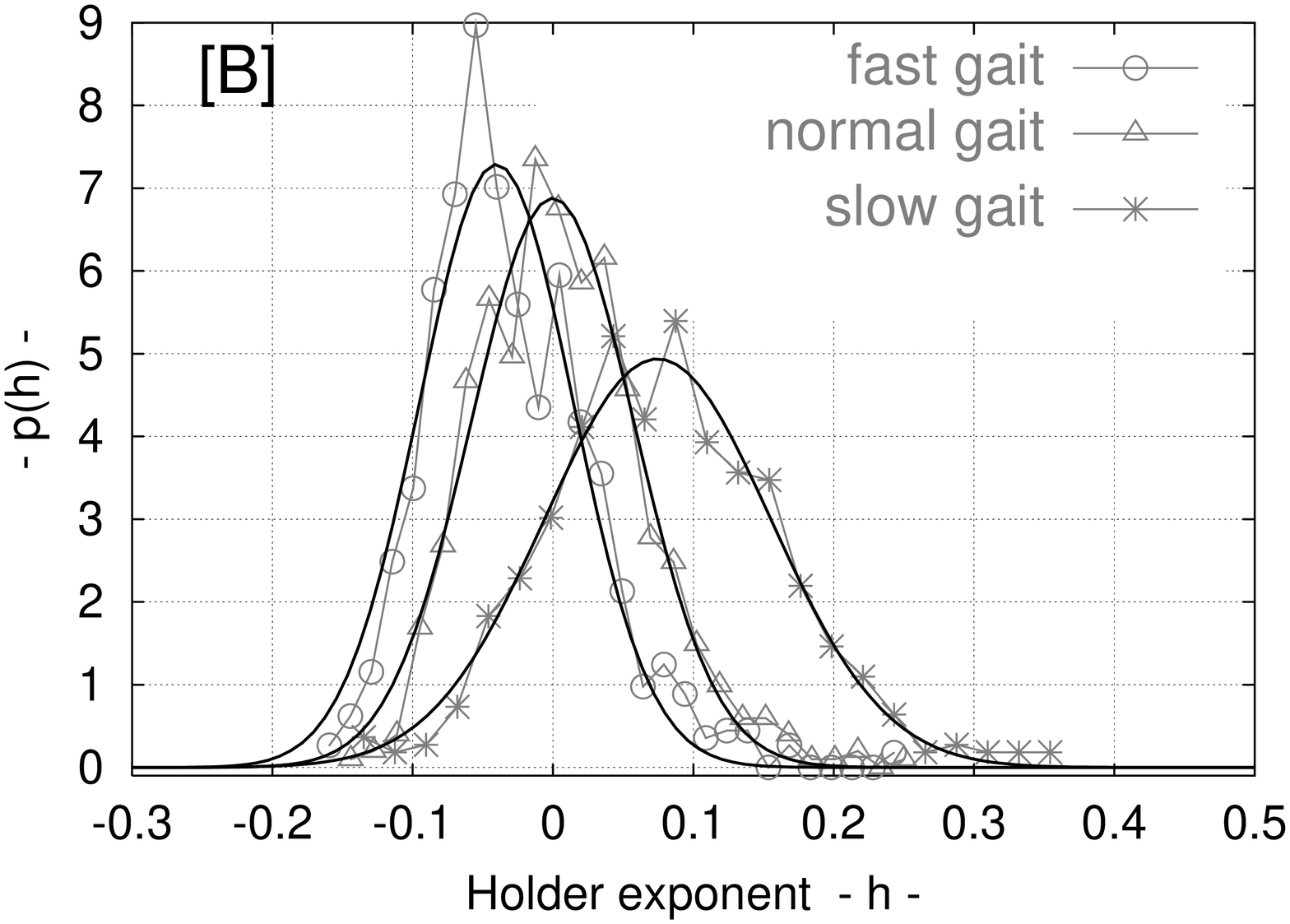,height=12cm,width=8cm,angle=-90}
\caption{ Histogram and probability density estimation of the H\"older exponents are shown for  walker number 5 (A) and  walker number 8 (B): slow-star, normal-triangle and fast-circle gait.  By changing the gait mode from slow to normal the mean Holder exponent $\overline{h}$ decreases but from normal to fast it usually increases,  but may also decrease. The fitting curves are Gaussian functions, the mean value $h_0$ and the standard deviation $\sigma$ are in Table 3.    }
\end{figure}

\newpage

\begin{figure}

Figure 9\\
\epsfig{file=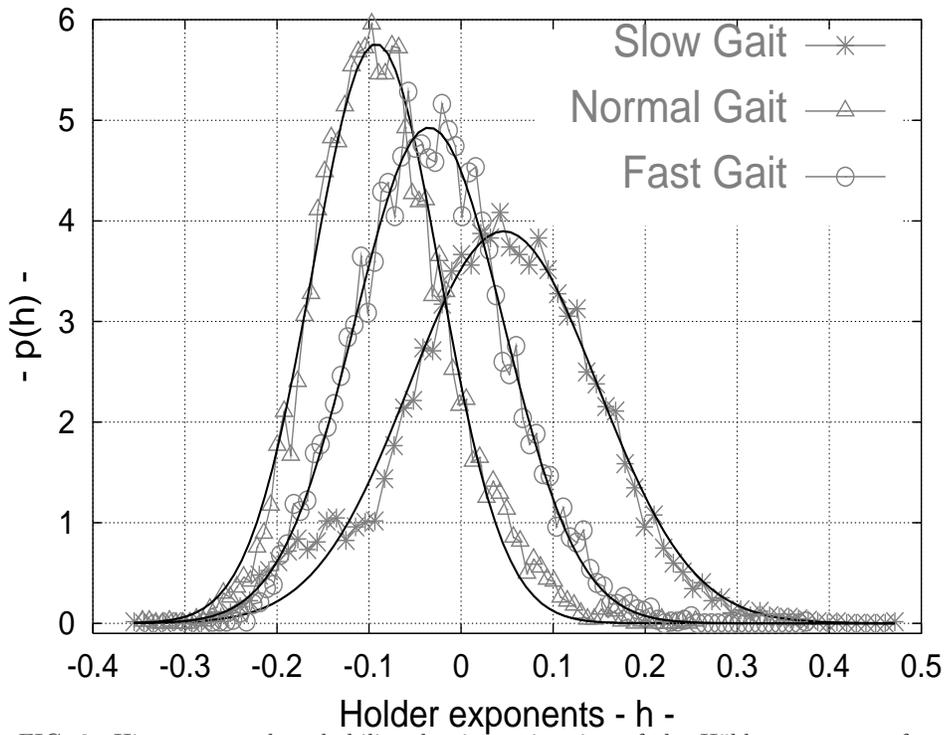,height=13cm,width=10cm,angle=-90}
\caption{Histogram and probability density estimation of the H\"older exponents for the three walking groups are shown: slow-star, normal-triangle and fast-circle gait.  By changing the gate mode from slow to normal the mean Holder exponent $h_0$ decreases but from normal to fast it increases. There is also an increasing of the width of the distribution $\sigma$ by moving from the normal to the slow or fast gait mode. The fitting curves are Gaussian functions, and the mean value $h_0$ and the standard deviation $\sigma$ are in Table 4.   }
\end{figure}

\newpage
\begin{figure}

Figure 10\\
\epsfig{file=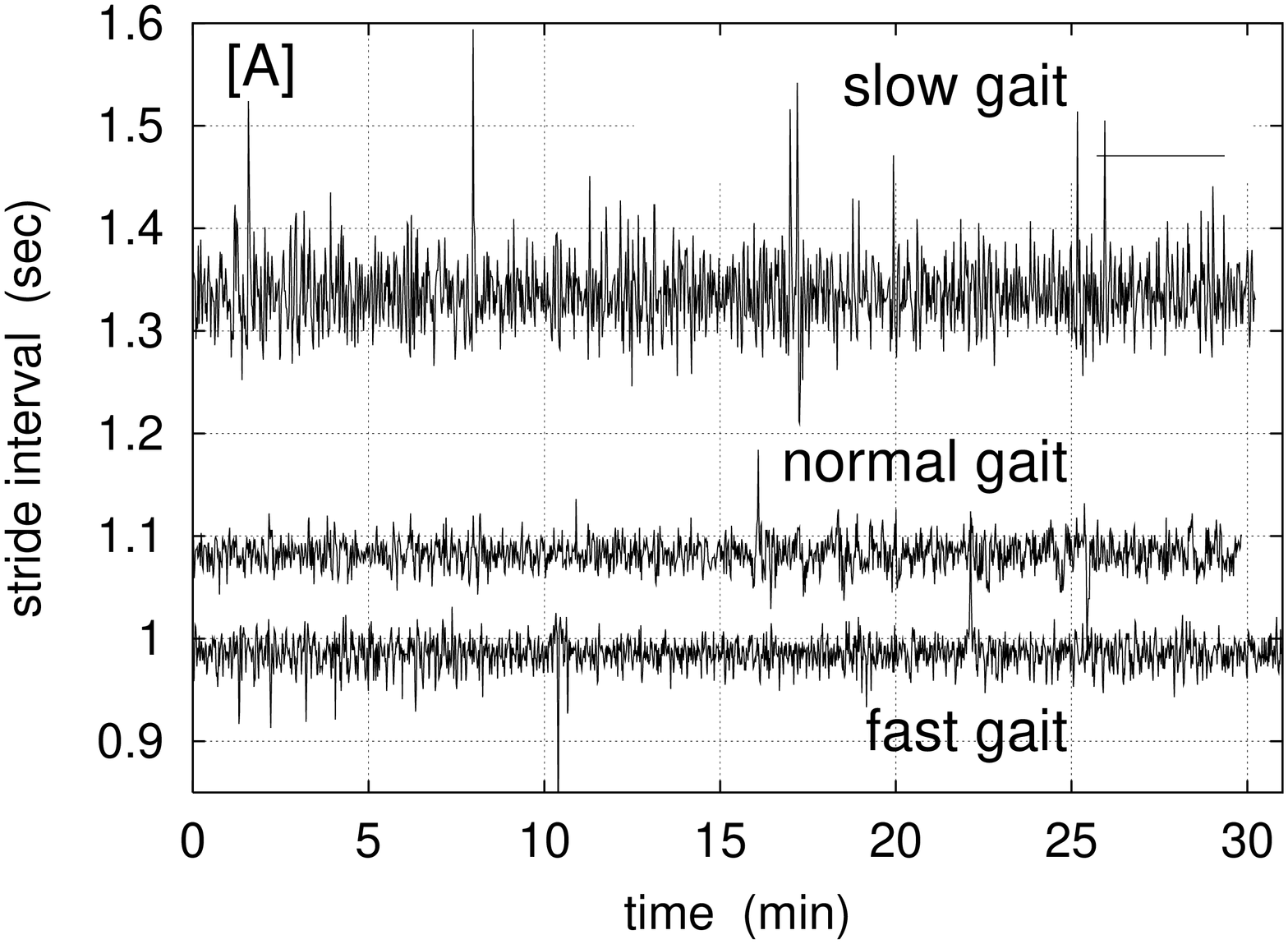,height=12cm,width=8cm,angle=-90}\\
\epsfig{file=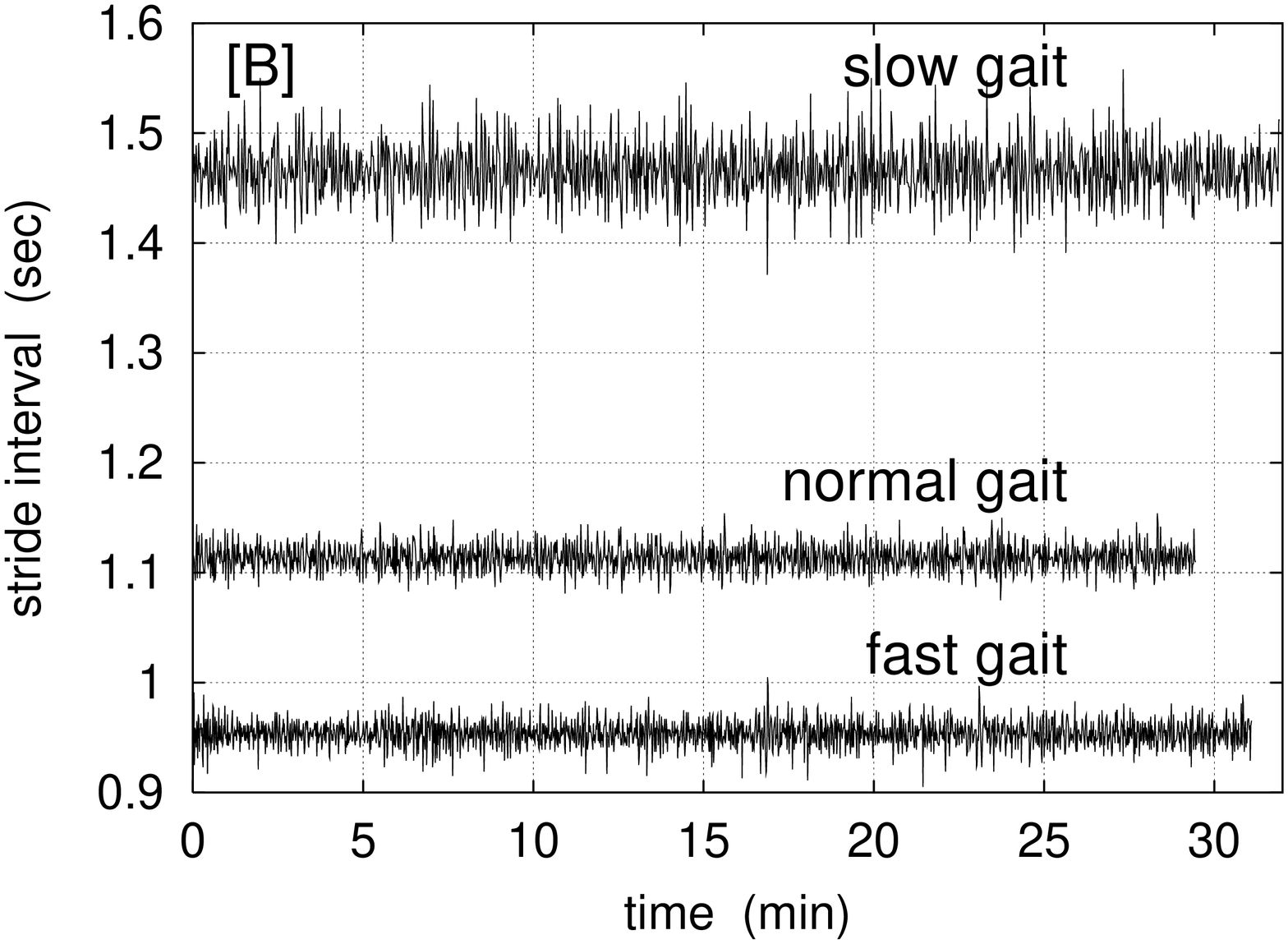,height=12cm,width=8cm,angle=-90}
\caption{  Stride intervals for slow, normal and fast gait for metronomically triggered walking is depicted. The total period of time is approximately 30 minutes. (A) person number 3; (B) person number 5.  }
\end{figure}

\newpage

\begin{figure}

Figure 11\\
\epsfig{file=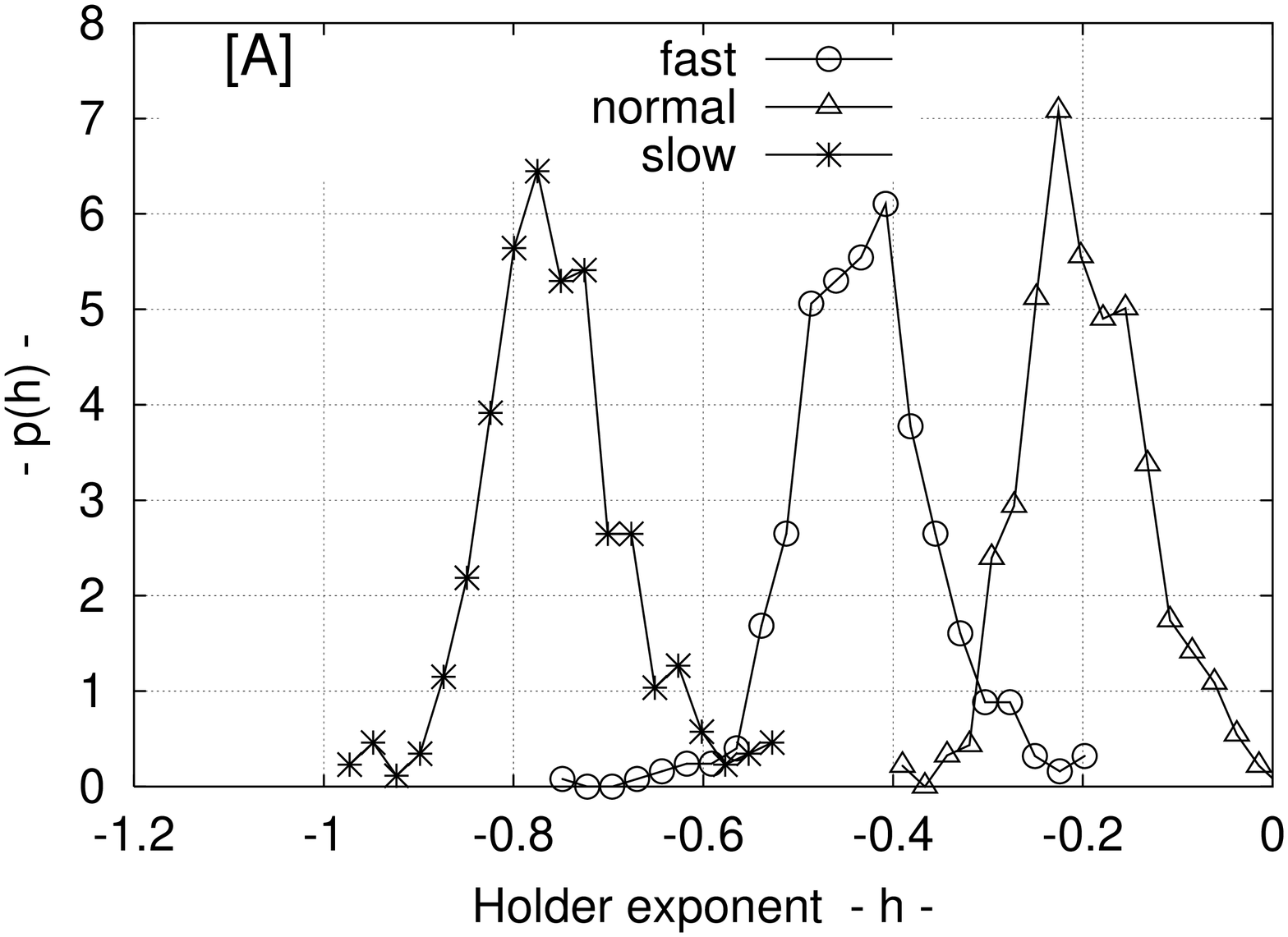,height=12cm,width=8cm,angle=-90}\\
\epsfig{file=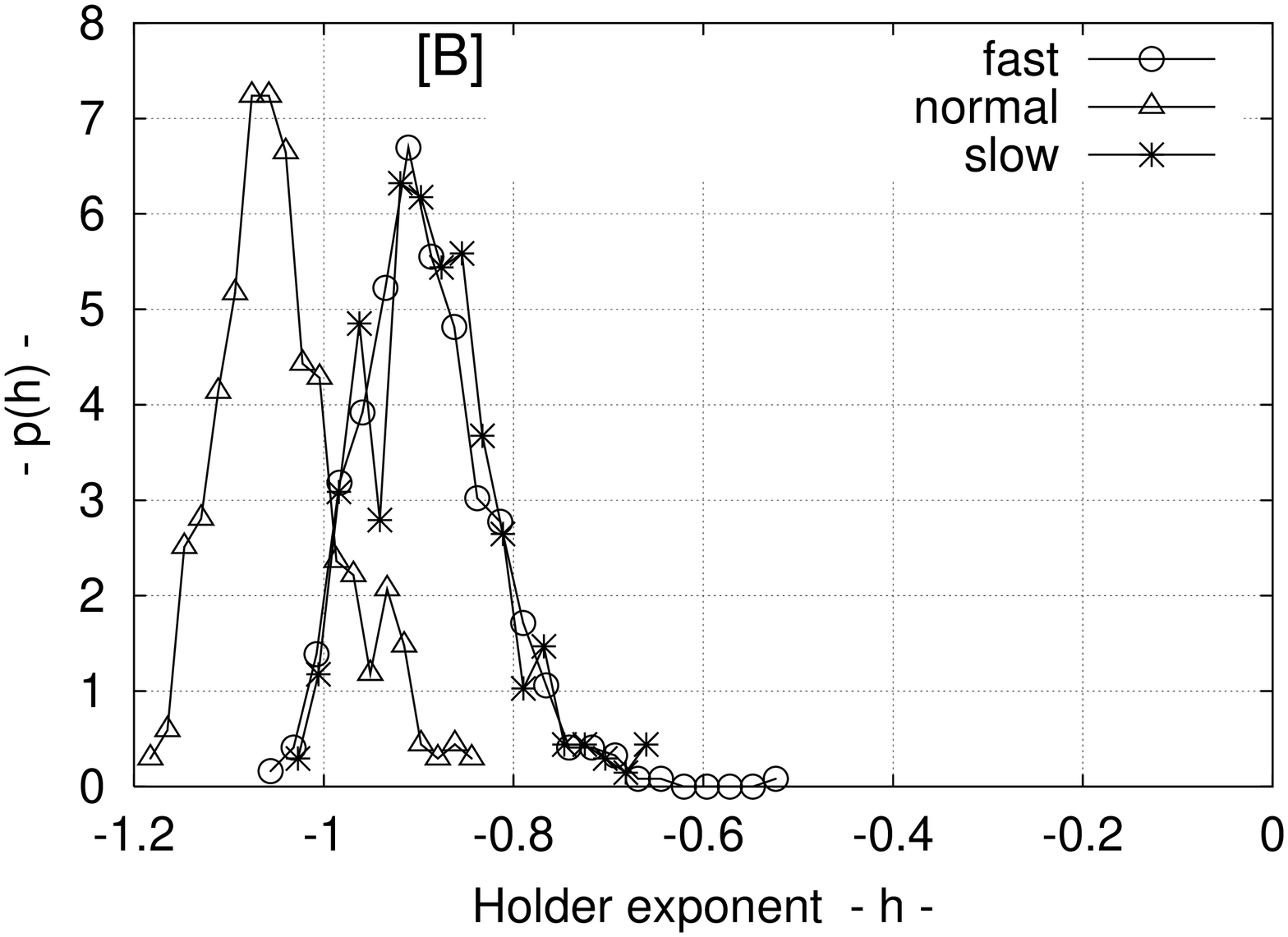,height=12cm,width=8cm,angle=-90}
\caption{ Metronomic walking. Histogram  estimation of the H\"older exponents for the walker number 3 (A), and the walker number 5 (B): slow-star, normal-triangle and fast-circle gait.    }
\end{figure}

\newpage

\begin{figure}

Figure 12\\
\epsfig{file=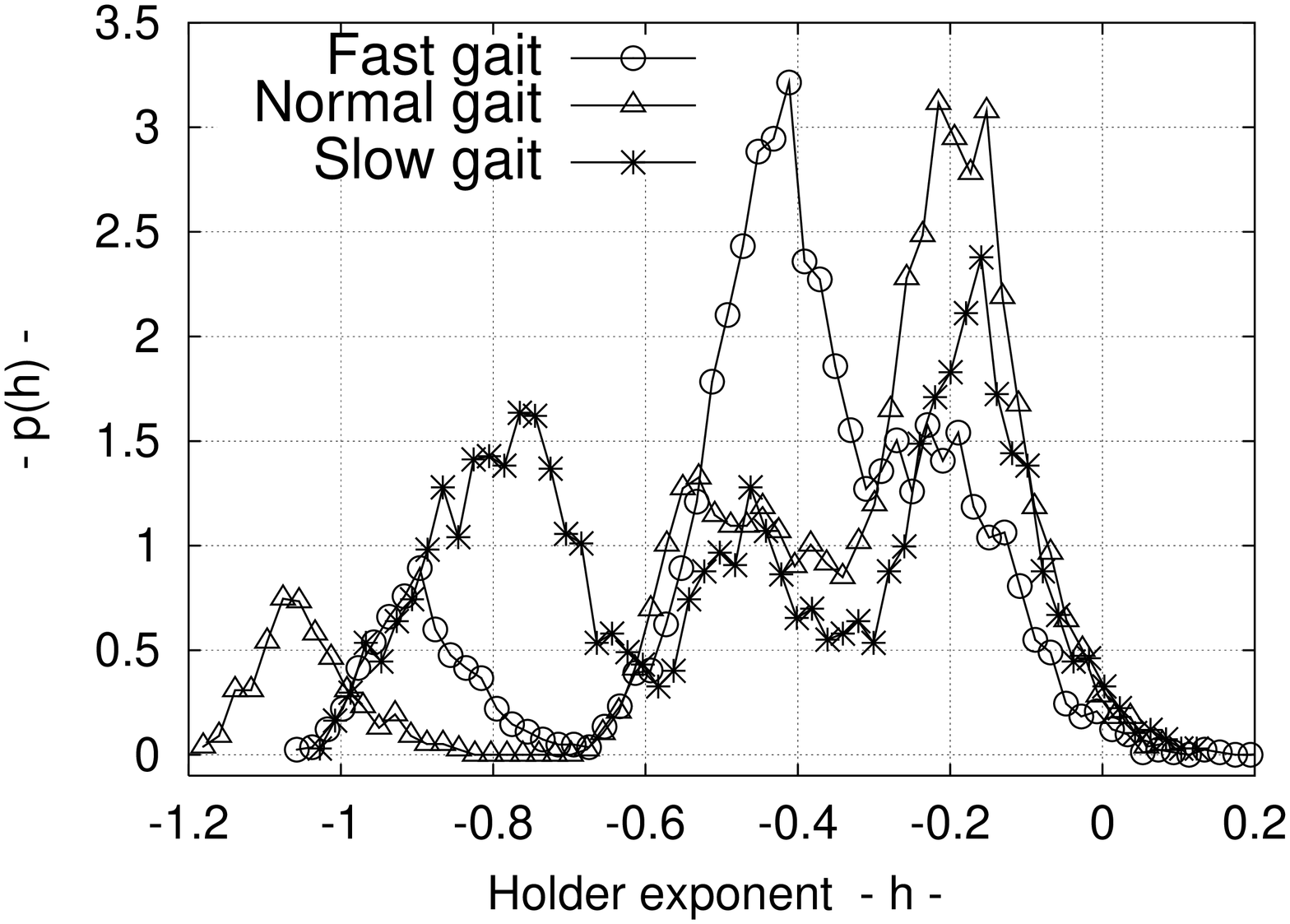,height=13cm,width=10cm,angle=-90}
\caption{Metronomic walking. Histogram estimation of the H\"older exponents for the three walking groups: slow-star, normal-triangle and fast-circle gait.  The data shows a large range of different conditions compatible with
 noise from antipersistent to persistent correlations.   }
\end{figure}


\begin{references}

\bibitem{bass94}  J.B. Bassigthwaighte, L.S. Liebowitch and B.J. West, {\it Fractal Physiology}, Oxford University Press, Oxford (1994).

\bibitem{struzik} Z. R. Struzik, Fractals, Vol. 8, No. {\bf 2},   163-179 (2000).

\bibitem{physionet} http://www.physionet.org/ .


\bibitem{Hausdorff4} J.M.  Hausdorff, P.L. Purdon, C.-K. Peng, Z. Ladin, J.Y. Wei, A.L. Goldberger, J. Appl. Physiol.  {\bf 80}, 1448-1457, (1996).  



\bibitem{collins1}  J.J. Collins and S.A. Richmond,  Biol. Cyb. {\bf 71}, 375-385 (1994).

\bibitem{mann}  M.D. Mann, {\it The Nervous System and Behavior}, Harper \& Row, Philadelphia (1981).

\bibitem{cohen}  A.H. Cohen, S. rossignol, and S. Grillner, Editors, {\it Neural control of rythmic movements in vertebrates}, Wiley, New York (1988).

\bibitem{collins2}  J.J. Collins and I.N. Stewart,  J. Nonlinear Sci. {\bf 3}, 349-392 (1993).

\bibitem{vierordt}  Vierordt, {\it Ueber das Gehen des Meschen in Gesunden und Kranken Zustaenden nach Selb-stregistrireden Methoden},  Tuebigen, Germany (1881).


\bibitem{hausdorff1}  J.M. Hausdorff, C.-K.Peng, Z. Ladin, J.Y. Wei and A.L.
Goldberger,  J. Appl. Physiol. {\bf 78}, 349-358 (1995).


\bibitem{west1}  B.J. West and L. Griffin,  Chaos, Solitions \& Fractals {\bf 10},
1519-1527 (1999).

\bibitem{west2}  B.J. West and L. Griffin,  Fractals {\it 6}, 101-108 (1998).

\bibitem{griffin}  L. Griffin, D.J. West and B.J. West,  J. Biol. Phys. {\bf 26}, 185-202 (2000).

\bibitem{ashkenazy} Y. Ashkenazy, J.M. Hausdorff, P. Ivanov, A.L. Goldberger and H.E. Stanley,  cond-mat/0103119 v1. 

\bibitem {bjwest}B.J. West, M. Latka and L. Griffin,  submitted to PRE.


\bibitem{2Mandelbrot}B.B. Mandelbrot, {\it The Fractal Geometry of Nature}, Freeman, New York, (1983).


\bibitem{daubechies} I. Daubechies, {\it Ten Lectures On Wavelets}, SIAM  (1992).

\bibitem{Mallat} S. G. Mallat, {\it A Wavelet Tour of Signal Processing} (2nd edition), Academic Press, Cambridge (1999). 


\bibitem{percival} D. B. Percival and A. T. Walden, {\it Wavelet Methods for Time Series Analysis}, Cambridge University Press, Cambridge (2000).

\bibitem{mallat2} S. G. Mallat, W. L. Hwang, IEEE Trans. on Information Theory {\bf 38}, 617-643 (1992).

\bibitem{mallat3} S. G. Mallat, S. Zhong, IEEE Trans. PAMI {\bf 14}, 710-732 (1992).

\bibitem{arneodo1} A. Arneodo, E. Bacry, J. F. Muzy, PRE {\bf 47}, No. 2, 875-884 (1993).

\bibitem{arneodo2} A. Arneodo, E. Bacry, J. F. Muzy, {\it International Journal of Bifurcation and Chaos} {\bf 4}, No. 2, 245-302 (1994).

\bibitem{Stanley1} P. C. Ivanov, M. G. Rosenblum, L. A. Nunes Amaral, Z. R. Struzik, S. Havlin, A. L. Goldberger and H. E. Stanley , {\it Nature} {\bf 399}, 461-465 (1999).

\bibitem{struzik2} Z. R. Struzik, CWI Report, INS-R9803 (1998).



\bibitem{feders} J. Feders, \emph{Fractals}, Plenum Publishers,
New York (1988).

\bibitem{beck} C. Beck, F. Schl\"ogl, {\it Thermodynamics of chaotic systems}, Cambridge university press , Cambridge, 1993. 



\bibitem{nr} W.H. Press, S.A. Teukolsky, W.T. Vetterling, B.P. Flannery, {\it Numerical Recipes in C}, Cambridge University Press, New York, 1997.   

\bibitem{nakamura} K. Nakamura, {\it Quantum Chaos}, Cambridge University Press, Cambridge (1993).

\end{references}
\end{document}